\documentclass[aps,pra,preprint,longbibliography]{revtex4-1}
\usepackage{amsfonts}
\usepackage{amssymb}
\usepackage{amsmath}
\usepackage{graphicx}
\usepackage[utf8]{inputenc}
\usepackage[format=plain,
      justification=RaggedRight,
      singlelinecheck=false]{caption}
\usepackage[justification=centering]{subcaption}

\usepackage{todonotes}

\usepackage{color}

\usepackage{graphicx}
\usepackage{latexsym}
\usepackage{amsmath}
\usepackage{amssymb}
\usepackage{bm}
\usepackage[colorlinks=true]{hyperref} 
\hypersetup{
    bookmarks=true,         
    unicode=false,          
    pdftoolbar=true,        
    pdfmenubar=true,        
    pdffitwindow=false,     
    pdfstartview={FitH},    
    pdftitle={Light induced enhancement of superconductivity
    via melting of competing bond-density wave order in underdoped cuprates},   
    pdfauthor={Aavishkar A. Patel and Andreas Eberlein},     
    pdfnewwindow=true,      
    colorlinks=true,       
    linkcolor=magenta, 
    citecolor=blue,        
    filecolor=magenta,      
    urlcolor=cyan           
}

\begin{document}

\title{Light induced enhancement of superconductivity
via melting of competing bond-density wave order in underdoped cuprates}

\date{\today}

\author{Aavishkar A. Patel}
\affiliation{Department of Physics, Harvard University, Cambridge MA 02138, USA}

\author{Andreas Eberlein}
\affiliation{Department of Physics, Harvard University, Cambridge MA 02138, USA}



\newcommand{\nn}{\nonumber \\}
\newcommand{\bx}{{\bm x}}
\newcommand{\br}{{\bm r}}
\newcommand{\bs}{{\bm s}}
\newcommand{\bt}{{\bm t}}
\newcommand{\bk}{{\bm k}}
\newcommand{\bp}{{\bm p}}
\newcommand{\bl}{{\bm l}}
\newcommand{\bq}{{\bm q}}
\newcommand{\bK}{{\bm K}}
\newcommand{\bQ}{{\bm Q}}

\newcommand{\etal}{\textit{et~al.}}
\newcommand{\ie}{\textit{i.\,e.}}


\hyphenation{counter-term}

\begin{abstract}
We develop a theory for light-induced superconductivity in underdoped cuprates 
in which the competing bond-density wave order is suppressed by driving phonons 
with light. Close to a bond-density wave instability in a system with a small 
Fermi surface, such as a fractionalized Fermi liquid, we show that the coupling 
of electrons to phonons is strongly enhanced at the bond-density wave ordering 
wavevectors, leading to a strong softening of phonons at these wavevectors. For 
a model of classical phonons with anharmonic couplings, we show that the 
combination of strong softening and driving can lead to large phonon 
oscillations. When coupled to a phenomenological model describing the 
competition between bond-density wave order and superconductivity, these phonon 
oscillations melt bond-density wave order, thereby enhancing pairing 
correlations.
\end{abstract}

\maketitle

\section{Introduction}
\label{sec:intro}
Superconductivity in cuprate materials still poses many questions that lack a 
theoretical understanding. One is about the maximum value of the critical 
temperature for superconductivity ($T_c$) that could be achieved due to the 
pairing mechanism, which also concerns the effects that limit $T_c$ in 
practice. Traditionally, the answer to this question was refined by synthesizing 
new materials or by changing control parameters such as external pressure. A 
new approach is the manipulation of materials properties such as order 
parameters by driving phonons with light~\cite{Tobey2008,Foerst2011}.

In the underdoped regime, $T_c$ is presumably limited by the occurrence of a 
pseudogap. The nature of this pseudogap state has not yet been clarified, and 
it is often interpreted as a state with competing orders or preformed pairs. 
Recent experiments in which the properties of underdoped cuprates are 
manipulated by driving phonons with light are promising to provide further 
insights about the physics in the underdoped regime. A series of remarkable 
experiments reported for example evidence for transient superconducting states 
induced by light in underdoped Yttrium~\cite{Kaiser2014} (YBCO) and 
Lanthanum~\cite{Fausti2011} (LCO) based cuprates far above $T_c$, and even up 
to room temperature in the former case.

In Lanthanum-based cuprates, a transient superconducting state can be induced 
above $T_c$ around 12\% hole doping~\cite{Fausti2011,Casandruc2015,Hunt2015}, 
where superconductivity competes strongly with stripe order. In equilibrium and 
above $T_c$, stripe order frustrates hopping perpendicular to the copper oxygen 
layers and destroys interlayer coherence. Stimulation with light leads to a 
melting of stripe order, which enhances the interlayer coherence and gives 
rise to a transient superconducting state~\cite{Foerst2014a}.

The situation for Yttrium based compounds~\cite{Kaiser2014} is less clear. 
From an experimental point of view, there is evidence for the melting of a 
competing order parameter~\cite{Foerst2014,WHu2014}, redistribution of 
interlayer coupling~\cite{WHu2014} and non-equilibrium lattice 
distortions~\cite{Mankowsky2014,Mankowsky2015}. Evidence also exists for a 
precursor superconducting state in equilibrium~\cite{Dubroka2011}. Several 
theories try to explain the transient superconducting state as a product of a 
redistribution of spectral weight~\cite{Hoeppner2015} or parametric 
cooling~\cite{Denny2015} of phase fluctuations due to periodic driving, 
exploiting the bilayer structure of YBCO. An alternative scenario invokes 
competition between bond-density wave (BDW) order and superconductivity in a 
three-dimensional model, where light modulates the interlayer hopping and 
influences the competition of orders~\cite{Raines2015}.

In the experiments on YBCO, the enhancement of superconductivity is 
observed when mid-infrared light couples resonantly to infrared 
phonons~\cite{Kaiser2014}, which in turn excite Raman phonons via nonlinear 
couplings~\cite{Mankowsky2014,Subedi2014,Mankowsky2015}. It thus does not seem 
to be related to a photo-induced redistribution of 
quasiparticles~\cite{Eliashberg1970,Ivlev1971}. The phonon frequencies are 
mismatched with the plasma resonance and a direct coupling to phase fluctuations 
seems inefficient. All three above-mentioned theories have difficulties in 
explaining why BDW correlations melt and pairing correlations appear on roughly 
the same time scales, while the disappearance of pairing correlations and the 
reappearance of BDW correlations happens on very different 
scales~\cite{Foerst2014}. Moreover, the proposed mechanisms either make use of 
the fact that YBCO is a bilayer cuprate or the electron hopping perpendicular to 
the copper oxygen layers plays a prominent role.

It is generally believed that the important physics of cuprate superconductors 
takes place in the copper oxygen planes, with the spacer layers mainly serving 
as charge reservoirs. It is therefore interesting to ask whether light-induced 
superconductivity could be achieved in a single copper oxygen plane. This 
question is intriguing because such a mechanism could also be at work in other 
cuprate materials that have one or multiple copper oxygen layers per unit cell 
and where experiments are more difficult.

In this paper we propose a mechanism for light-induced superconductivity based 
on the competition between BDW correlations and superconductivity, which may be 
applicable slightly above the equilibrium $T_c$ where the system is close to a 
BDW instability~\cite{Ghiringhelli2012,Chang2012,Blanco-Canosa2013}. We describe 
the interplay of BDW and pairing fluctuations with a phenomenological non-linear 
sigma model~\cite{Efetov2013,Hayward2014}, and investigate the external driving 
of this model by coupling it to phonons via electrons. Parameter quenches of 
the non-linear sigma model have been studied by Fu~\etal\ in 
Ref.~\onlinecite{WFu2014}. The electrons are described as a fractionalized Fermi 
liquid (FL$^\ast$)~\cite{Senthil2003,Senthil2004}, which is a model for the 
pseudogap state in underdoped cuprates. It shows BDW correlations at an axial 
wave vector that connects the tips of Fermi arcs~\cite{Chowdhury2014a}, as seen 
in experiment~\cite{Comin2014}.

An important ingredient of our theory is the enhanced coupling of electrons and 
phonons at the BDW wave vector that is observed in experiments. It leads to a 
strong softening of the phonon dispersion~\cite{Raichle2011,LeTacon2014} and a
phonon linewidth that increases strongly with decreasing 
temperature~\cite{Blackburn2013a,LeTacon2014}. Our theory traces back the 
enhanced electron-phonon coupling to vertex corrections due to short-range 
antiferromagnetic fluctuations and explains the phonon renormalization above 
$T_c$ as being a consequence of strong BDW fluctuations. However, the enhanced 
electron-phonon coupling does not directly lead to an enhancement of $T_c$ 
because it is restricted to a small region in momentum space and certain 
non-linear couplings between phonons and pairing fluctuations nearly cancel in 
our model. Instead, the combination of strong phonon softening and phonon 
anharmonicities allows for the efficient driving of phonons near the BDW 
wavevector by light pulses. These strongly driven phonons suppress BDW 
fluctuations by enhancing their mass, thereby enlarging the relative size 
of the phase space for pairing, which enhances superconducting correlations.

\begin{figure}
	\centering
	\includegraphics[width=6.1in]{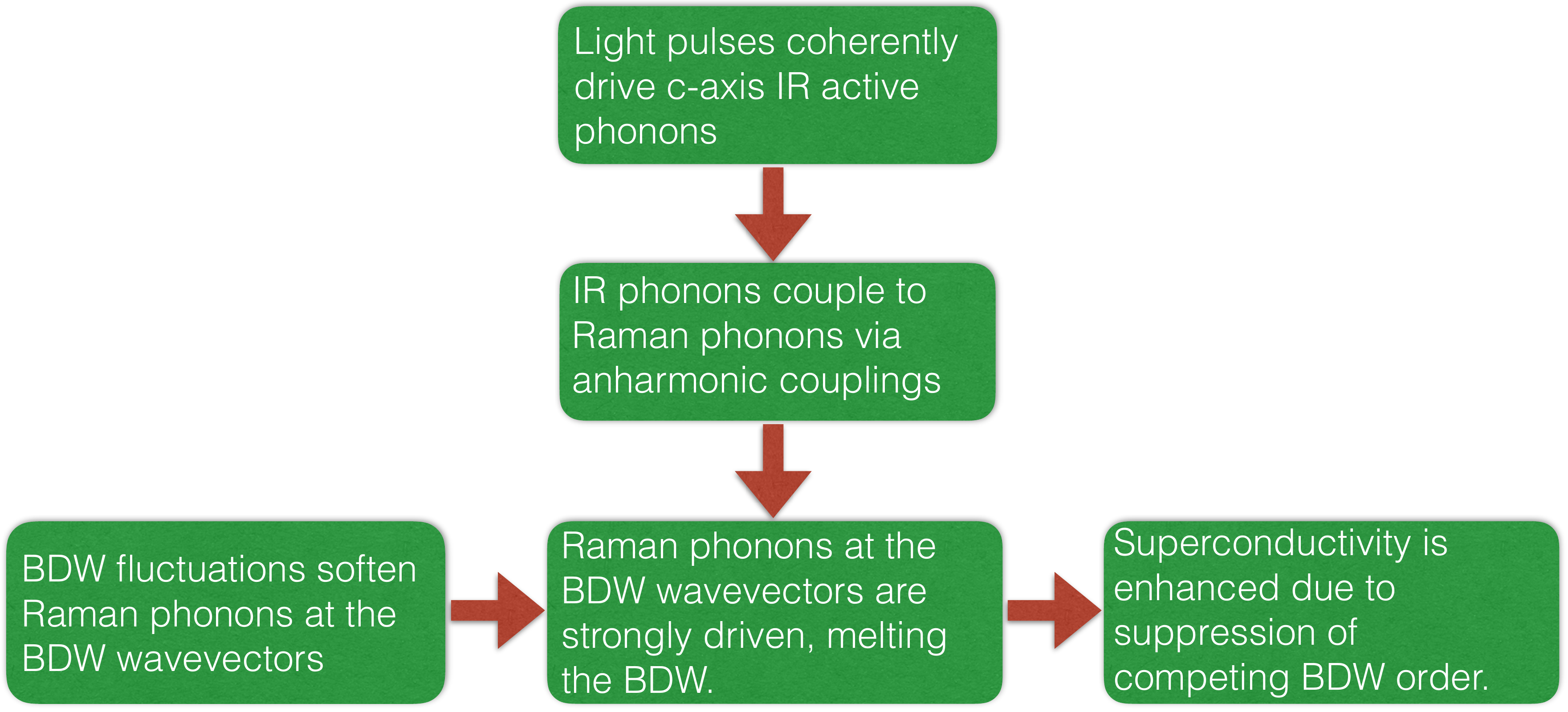}
	\caption{(Color online) Flow diagram illustrating how the interaction between 
light, driven phonons, electrons and order parameters can enhance
superconductivity.}
	\label{fig:flowdiagram}
\end{figure}
This paper is organized as follows. In Sec.~\ref{sec:model}, we introduce the 
effective models for competing BDW and pairing correlations, for electrons in 
underdoped cuprates (the FL$^\ast$), their mutual coupling as well as the 
coupling between electrons and phonons. The most important roles in our theory 
are played by order parameters and phonons, whereas the electrons only act as 
intermediaries between them. In Sec.~\ref{sec:phonon}, we introduce a model for 
driven phonons and describe its renormalization by interaction effects as well 
as its dynamics close to a BDW instability. We find strong phonon softening that 
can entail momentum-selective driving and large amplitudes at the BDW wave 
vector. In Sec.~\ref{sec:box}, we derive effective couplings between phonons, 
BDW and pairing fluctuations and describe how the driving of phonons impacts the 
interplay between BDW and pairing fluctuations. Large phonon amplitudes at the 
BDW wave vector can lead to a suppression of BDW correlations in favor of 
superconductivity. This interplay is summarized in Fig.~\ref{fig:flowdiagram}. 
In Sec.~\ref{sec:disc} we discuss our results and draw conclusions.

\section{Model}
\label{sec:model}
We are interested in the temperature region above the critical temperature 
for superconductivity $T_c$, where YBCO shows strong bond-density wave 
fluctuations~\cite{Ghiringhelli2012}. In this regime, the interplay between 
superconducting and bond-density wave fluctuations can be described by a 
Landau theory of competing orders. In order to minimize the number of couplings
in such a theory we choose to use a classical non-linear sigma model 
(NLSM) that correctly captures various qualitative features of the behavior of
competing orders in the pseudogap phase slightly above 
$T_c$~\cite{Efetov2013,Hayward2014}, 
\begin{gather}
\mathcal{Z}_{\rm cl} = \int \mathcal{D}\Psi\mathcal{D}\Psi^\ast\prod_{n=x,y} 
\mathcal{D}\Phi_n\mathcal{D}\Phi^\ast_n e^{-\beta S_{\rm 
cl}}~\delta(|\Psi|^2+\sum_{n=x,y}|\Phi_n|^2-1), \\
S_{\rm cl} = \rho_s \int 
d^2\mathbf{x}\left[\frac{1}{2}|\nabla\Psi|^2+\sum_{n=x,y}\left(\frac{\eta}{2}
|\nabla \Phi_n|^2+\frac{m^2}{2}|\Phi_n|^2\right)\right],
\end{gather}
where $\Psi$ and $\Phi_n$ are complex bosonic fields describing pairing and BDW 
fluctuations, respectively. They are combined into an $O(6)$ order parameter 
with fixed length in order to focus on angular fluctuations. At the lowest 
temperatures, superconductivity is favored due to the positive mass term for 
BDW fluctuations. As $T$ is increased, superconducting order is lost, which is 
aided by the strengthening of BDW fluctuations. As will become clear in the 
following sections, our mechanism does not specifically require this model and 
will work for a generic Landau theory of competing orders, since it relies only 
on the discrepancies in mass renormalizations of different order parameters. 

\begin{figure}
\centering
\begin{subfigure}[c]{3.1in}
		\includegraphics[width=3.0in]{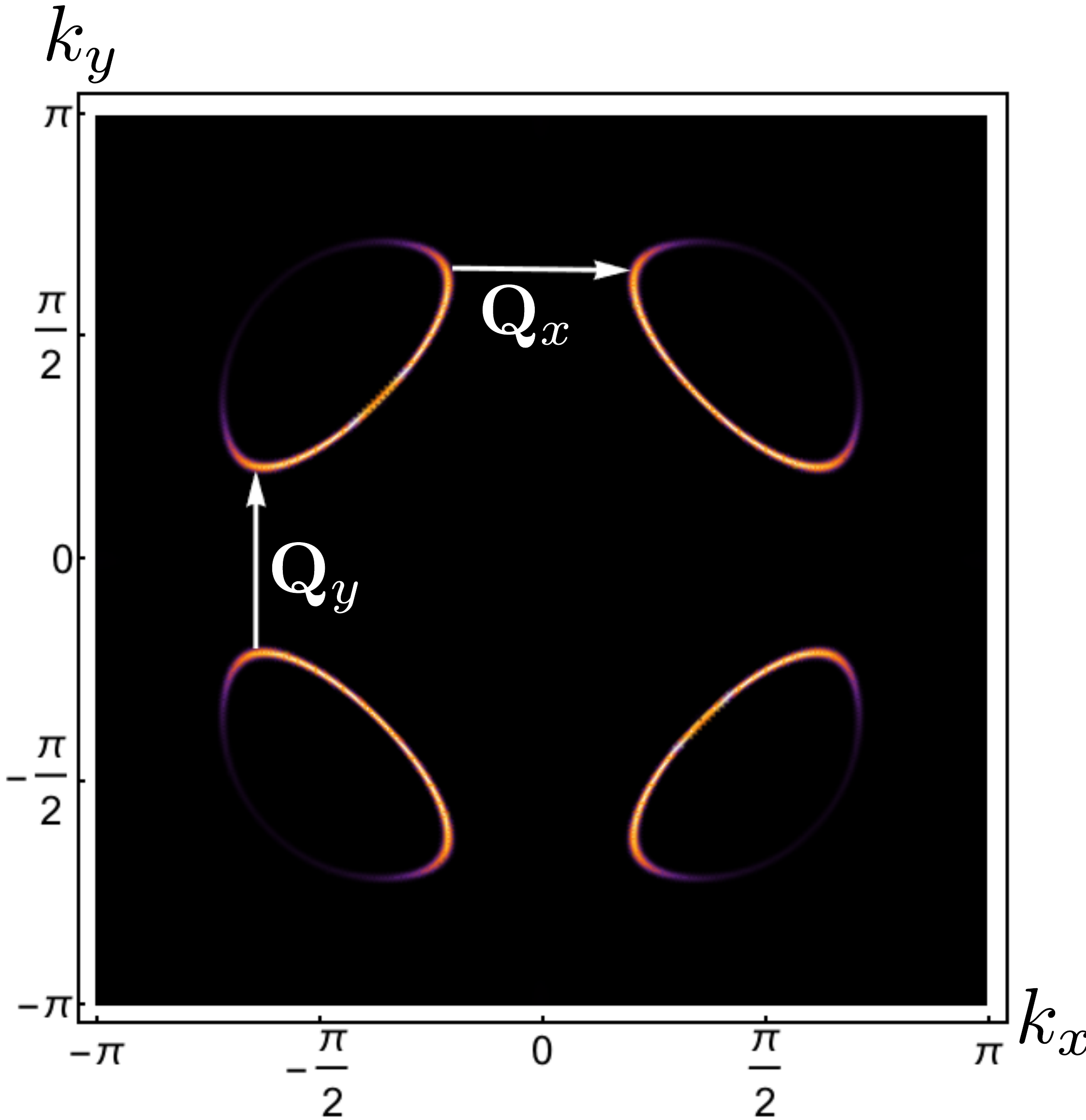}
		\caption{} \label{fig:FSsmall}
	\end{subfigure}
	\vspace{0.25in}
\begin{subfigure}[c]{3.1in}
		\includegraphics[width=3.0in]{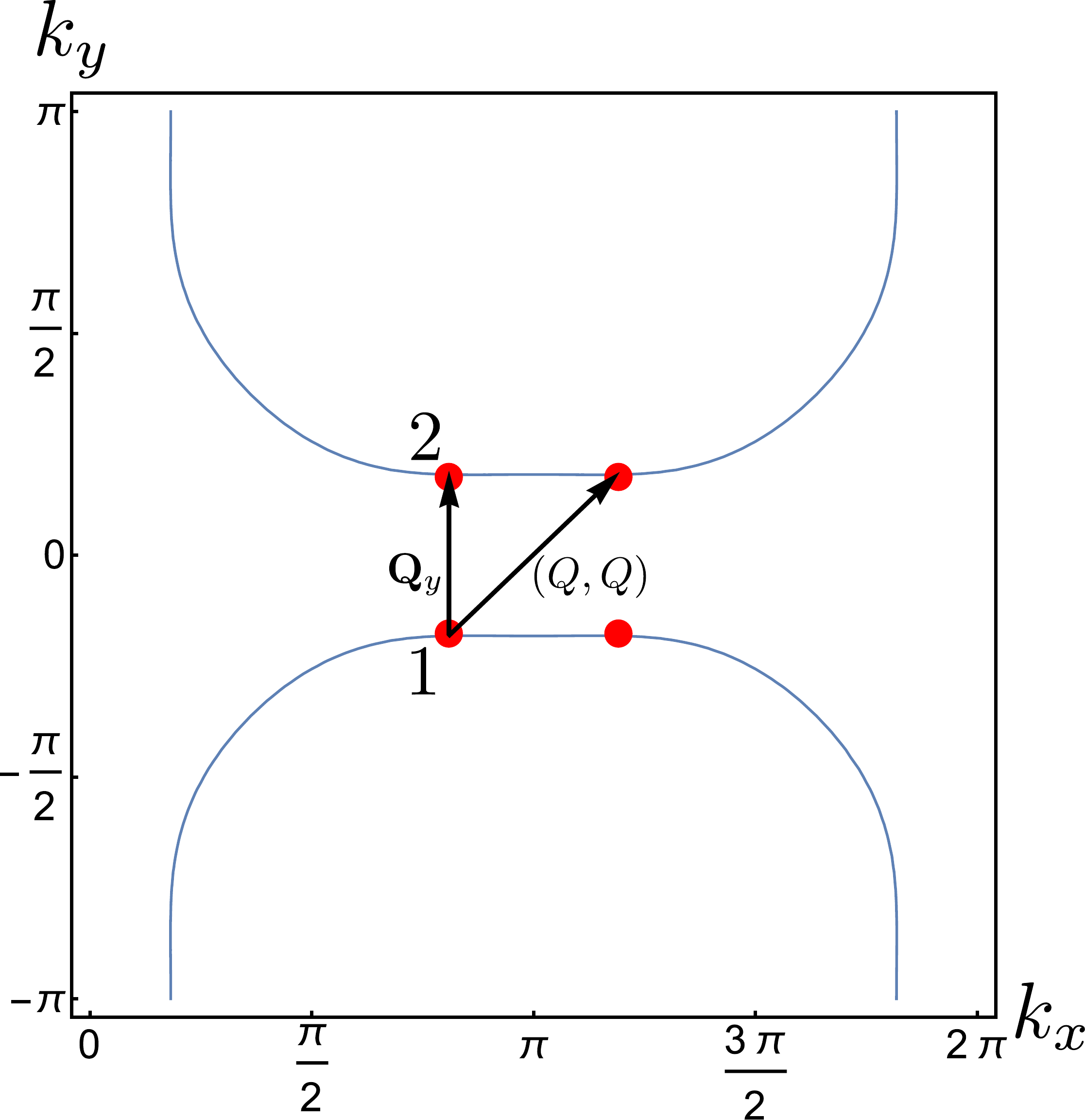}
		\caption{} \label{fig:FSlarge}
	\end{subfigure}
\caption{(Color online) Comparison between (a) small and (b) large Fermi 
surfaces. Also shown are possible BDW ordering wave vectors. These connect the 
tips of the hole pockets in (a) or the antiferromagnetic hot spots (marked in 
red) in (b). In (a) we show the spectral function at $\omega = 0$ with color 
encoding spectral weight. For this figure, and for the rest of this work, we 
use $t_1 = 1.0$, $t_2 = -0.32$, $t_3 = 0.128$, $\mu = -1.119$, 
$\tilde{t}_0=-0.5t_1$, $\tilde{t}_1=0.4t_1$, $\lambda=0.6t_1$.}
\label{fig:FS}
\end{figure}

We derive the coupling of the non-linear sigma model to phonons from a 
phenomenological model for electrons in underdoped cuprates, which we take as a 
fractionalized Fermi liquid (FL$^\ast$)~\cite{Senthil2003,Senthil2004}. In this 
model, the Fermi surface is gapped out in the anti-nodal region of the Brillouin 
zone and reconstructed into hole-like pockets with low spectral weight on their 
back-sides. It shows incommensurate BDW fluctuations that are peaked at axial 
wave vectors close to those seen in experiment~\cite{YQi2010, Chowdhury2014a, 
DCSS15Review}. The model is defined by the action 
\begin{gather}
\mathcal{L}_\text{f} = \sum_{k, \sigma} \psi^\dagger_{\sigma}(k) 
G^{-1}(k) 
\psi_\sigma(k) + \mathcal L_\text{int,f} + \mathcal L_\text{f-BDW} + 
\mathcal{L}_\text{f-dSC} + \mathcal{L}_\text{f-ph}
\end{gather}
where $\psi^{(\dagger)}$ are fermionic Grassmann fields, $k = (\omega_n, 
\mathbf{k})$ collects Matsubara frequencies $\omega_n$ and momenta 
$\mathbf{k}$, $\sigma = \uparrow$, $\downarrow$ and $G^{-1}$ is the inverse 
electron propagator. $\mathcal L_\text{int,f}$ describes short-range 
electron-electron interactions. The other terms couple electrons to BDW 
fluctuations, $d$-wave pairing fluctuations and phonons, respectively, and 
are detailed below. The electron propagator is given by
\begin{equation}
	G(k) = \frac{Z}{i\omega_n-\xi^+_{\mathbf{k}} - \Sigma(k)}
\label{eq:propagator}
\end{equation}
where
\begin{align}
	\xi^+_{\mathbf{k}} &= 
\varepsilon_\mathbf{k} +\tilde{\varepsilon}_\mathbf{k},\qquad\qquad\qquad 
\xi^-_\mathbf{k} = \varepsilon_\mathbf{k} -\tilde{\varepsilon}_\mathbf{k},\\
	\varepsilon_\mathbf{k} &= 
-2t_1(\cos k_x + \cos k_y) - 4 t_2 \cos k_x \cos k_y  - 
2 t_3 (\cos(2k_x)+\cos(2k_y))- \mu, \\
	\tilde{\varepsilon}_\mathbf{k} &= -\tilde{t}_0 - \tilde{t}_1 
(\cos k_x +\cos k_y)
\end{align}
is the electron dispersion,
\begin{equation}
	\Sigma(k) = \frac{\lambda^2}{i\omega_n-\xi^-_{\mathbf{k} + \mathbf{K}}}
\end{equation}
the self-energy of the FL$^\ast$ and $\mathbf{K}=(\pi,\pi)$. The hopping 
parameters used in this work are listed in the caption of Fig.~\ref{fig:FS}. 
The Fermi liquid form of the propagator in Eq.~\eqref{eq:propagator} is 
manifest if it is rewritten in the following ``two-band" form, 
\begin{gather}
G(k) = 
\sum_{\alpha = \pm}\frac{Z_\mathbf{k}^\alpha}{i\omega_n-E_\mathbf{k}^\alpha}, 
\qquad \qquad Z_\mathbf{k}^\alpha = 
\frac{Z^2(E_\mathbf{k}^\alpha - \xi_{\mathbf{k}+\mathbf{K}}^-)^2}{
\lambda^2+(E_\mathbf{k}^\alpha - \xi_{\mathbf{k}+\mathbf{K}}^-)^2}, \\
E_\mathbf{k}^\pm = \frac{1}{2}(\xi_{\mathbf{k}}^+ + 
\xi_{\mathbf{k}+\mathbf{K}}^-)\pm\sqrt{\lambda^2+\frac{1}{4}(\xi_{\mathbf{k}}^+ 
- \xi_{\mathbf{k}+\mathbf{K}}^-)^2}.
\end{gather}
The gap parameter $\lambda$ arises from short-range antiferromagnetic 
order~\cite{YQi2010} and leads to the reconstruction of the Fermi surface. The 
parameters $\tilde t_0$ and $\tilde t_1$ control the location of the hole 
pockets. For $\lambda \neq 0$ but $\tilde t_0 = \tilde t_1 = 0$, they are 
centered at $(\pm \pi/2, \pm \pi/2)$. Setting these FL$^\ast$ parameters to 
zero, we recover the ``large Fermi surface". Examples for a large and small 
Fermi surface are shown in Fig.~\ref{fig:FS}. We will also comment on the 
impact of using a small vs. a large Fermi surface on our theory. The parameter 
$Z$ measures the weight of the fermionic quasiparticles relative to that of 
incoherent excitations and is not important for our purposes. We thus set $Z=1$.

We take into account nearest-neighbor exchange interactions,
\begin{equation}
\mathcal{L}_\text{int,f} = 
 \sum_{k, k', q, \sigma_i} 
J(\mathbf{q}) \boldsymbol \tau_{\sigma_1\sigma_4}
\cdot\boldsymbol\tau_{\sigma_2 \sigma_3} \psi^\dagger_{k + q /2,\sigma_1} 
\psi^\dagger_{k' - q /2,\sigma_2} \psi_{k' + q/2, \sigma_3} \psi_{k - q/2, 
\sigma_4},
\end{equation}
where $J(\mathbf{q}) = 2 J (\cos q_x + \cos q_y)$, in order to mimic the 
strong short-range antiferromagnetic correlations that are found in underdoped 
cuprates. These can induce BDW instabilities~\cite{Chowdhury2014a} of the 
FL$^\ast$. For simplicity we restrict ourselves to nearest-neighbor exchange. 
Further neighbor and density-density interactions could be added, but are not 
expected to change our results in an essential way.

In order to study the interaction between phonons and pairing as well as BDW 
fluctuations, we introduce fermion-boson couplings. The coupling of electrons 
to BDW fluctuations and Cooper pairs could be derived from Hubbard-Stratonovich 
transformations of short-range interactions. For simplicity, we use 
phenomenological expressions and couple BDW fluctuations and electrons
via~\cite{Allais2014,Chowdhury2014a}
\begin{equation}
\mathcal{L}_{\rm f-BDW} = \lambda_\text{BDW} \sum_{k, q,n=\{x,y\},\sigma} 
g_{\mathbf{Q}_n}(k) 
\psi^\dagger_{k + \mathbf{Q}_n/2 + q/2,\sigma} \psi_{k-\mathbf{Q}_n/2 
- q/2,\sigma} \Phi_n(q)+ \mathrm{c.c.},
\end{equation}
where $\lambda_\text{BDW}$ is the coupling strength and $n = \{x, y\}$ sums 
over the wavevectors at which the BDW fluctuations are strongest, 
$\mathbf{Q}_x=(\pm q, 0)$ and $\mathbf{Q}_y=(0, \pm q)$. The intra-unit cell 
structure of the BDW fluctuations has predominantly $d$-wave 
character~\cite{Chowdhury2014a,YWang2014,Allais2014a,Comin2015}. We therefore 
approximate the form factor $g_{\mathbf{Q}_n}(\mathbf{k})$ by
\begin{equation}
g_{\mathbf{Q}_n}(\mathbf{k}) \approx g_d(\mathbf{k}) = \cos k_x - \cos k_y
\end{equation}
and neglect possible admixtures with $s$-wave symmetry for simplicity. $d$-wave 
pairing fluctuations couple to the electrons as
\begin{equation}
\mathcal{L}_{\rm f-dSC} = \lambda_\text{dSC} \sum_{\mathbf{k}, 
\mathbf{q},\sigma} g_d(\mathbf{k}) \psi^\dagger_{q/2 + k,\uparrow} 
\psi^\dagger_{q/2 - k,\downarrow} \Psi(q) + \mathrm{c.c.},
\end{equation}
where $\lambda_\text{dSC}$ is the coupling constant. The value of the coupling 
constants $\lambda_\text{BDW}$ and $\lambda_\text{dSC}$ is not important and 
expected to be of $\mathcal O(1)$.

The coupling of electrons to phonons, which are described by real scalar fields 
$\varphi$, is modeled by
\begin{gather}
\mathcal{L}_{\rm f-ph} = \lambda_\text{ph}\sum_{k, q, \sigma} g_{\rm 
ph}(\mathbf{q},\mathbf{k})  \psi^\dagger_{k + 
q/2,\sigma} 
\psi_{k-q/2,\sigma}\varphi(q) +  \mathrm{c.c.}~.
\end{gather}
In the following we consider the coupling of electrons to buckling modes of the 
Copper oxygen plane, for which we can generically consider $g_{\rm 
ph}(\mathbf{q},\mathbf{k}) = g_{\rm ph}(\mathbf{q})$, 
with~\cite{Bulut1996}
\begin{equation}
|g_{\rm ph}(\mathbf{q})|^2 = \cos^2\left(\frac{q_x}{2}\right) + 
\cos^2\left(\frac{q_y}{2}\right).
\end{equation}
The dependence on $\mathbf{q}$ is not very important and our conclusions do 
not change in case a momentum independent bare vertex, $g_{\rm ph}(\mathbf{q}) 
= g_\text{ph}$, is used. In the next section, we will see that strong 
fluctuations at an incipient BDW instability will induce a $\mathbf{k}$ 
dependence for the electron-phonon vertex. The action describing the phonon 
dynamics is also detailed in the next section.

\section{Phonon renormalization and phonon dynamics near bond-density wave 
instability}
\label{sec:phonon}
In the experiments on YBCO~\cite{Kaiser2014}, the laser pulses drive c-axis 
infrared (IR) phonons. Since the pulse wavelength is much larger than the 
lattice constants, the IR phonons can be considered to be excited at zero 
momentum. The IR phonons then couple to the in-plane Raman phonons, which in 
turn couple to the electrons~\cite{Subedi2014,Knap2015-arXiv}. We model the 
dynamics of the Raman phonons with the classical real-time Lagrangian with 
quartic
anharmonicities developed by Subedi~\etal\ for cuprate 
superconductors~\cite{Subedi2014}, 
\begin{equation}
\mathcal{L}_{\rm ph} = 
\int d^2\mathbf{x}~ \frac{1}{2}(\partial_t\varphi(\mathbf{x}))^2 - 
\frac{1}{2} \omega_{\rm ph, 0}^2 \varphi(\mathbf{x})^2 + 
\frac{r}{2} \phi_{\rm IR}(\mathbf{x})^2 \varphi(\mathbf{x})^2  - 
\frac{u}{4}\varphi(\mathbf{x})^4,
\label{eq:PhononAction}
\end{equation}
where $\omega_{\rm ph,0}$ is the bare Raman phonon frequency. The dispersion of 
the (optical) Raman phonons is mostly flat throughout the Brillouin zone, so we 
do not include gradient terms in the Lagrangian. $\phi_{\rm IR}(\mathbf{x})\sim 
A_{\rm IR}\cos(\omega t)$ is the amplitude of the driven IR phonon and $r$ the 
coupling between (odd under inversion) IR and (even under inversion) Raman 
phonons. Since the Lagrangian must be inversion symmetric, the Raman phonons 
must couple to $\phi_{\rm IR}^2$ instead of $\phi_{\rm IR}$ at lowest order. The 
anharmonic coupling $u>0$ is important in order to stabilize the strongly driven 
system. Note that the IR phonon does not possess any dynamics besides the 
driving and that all phonons are described as classical oscillators. This is 
justifiable for the strongly driven phonons in the experiments, where the 
phonon amplitudes reach a few percent of the lattice 
constants~\cite{Mankowsky2014,Mankowsky2015}.

Before studying the phonon dynamics described by the action in 
Eq.~\eqref{eq:PhononAction}, we compute the phonon renormalization near a BDW 
instability. In such a situation, a strong renormalization of phonons has been 
observed in x-ray scattering experiments~\cite{Blackburn2013a, LeTacon2014}. In 
the experiment by Le~Tacon~\etal~\cite{LeTacon2014}, a significant softening of 
the dispersion of IR phonons together with a strongly increased linewidth is 
observed at the BDW wave vector. As described by our model below, this effect 
should carry over to Raman phonons as well.

Intuitively, we expect that the short-range antiferromagnetic fluctuations that 
drive the BDW instability at a particular wavevector can also significantly 
renormalize the electron-phonon vertex at that wavevector. We thus consider the 
renormalization of this vertex by short-range antiferromagnetic fluctuations 
close to a BDW instability via a Bethe-Salpeter equation.
\begin{table}
\centering
\begin{tabular}{|c|c|}
\hline
$\ell$ & Basis Function $f_\ell(\mathbf{k})$  \\ \hline
0      & 1                     \\ \hline
1      & $\cos(k_x) + \cos(k_y)$ \\ \hline
2      & $\cos(k_x) - \cos(k_y)$ \\ \hline
3      & $\sin(k_x) + \sin(k_y)$ \\ \hline
4      & $\sin(k_x) - \sin(k_y)$ \\ \hline
\end{tabular}
\caption{Basis functions for decomposing the renormalized electron-phonon 
vertex}
\label{tab:basis}
\end{table}
\begin{figure}
	\begin{subfigure}[c]{\linewidth}
		\includegraphics[scale=.9]{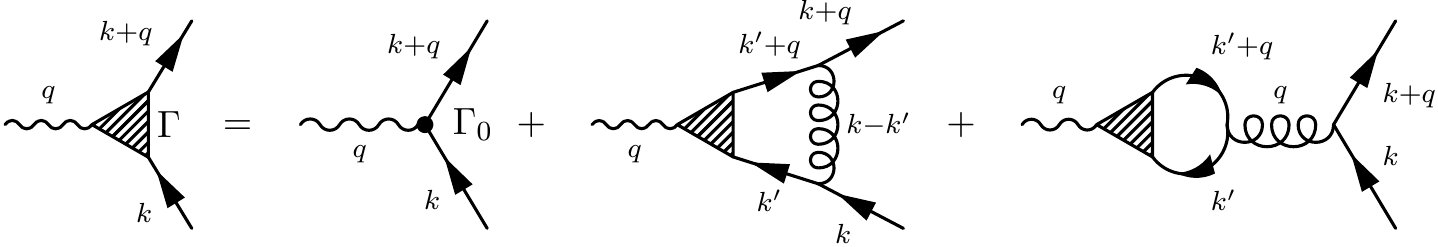}
		\caption{} \label{fig:BS_a}
	\end{subfigure}

	\begin{subfigure}[c]{\linewidth}
		\includegraphics[scale=.9]{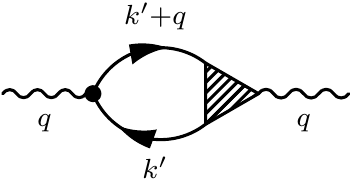}
		\caption{} \label{fig:BS_b}
	\end{subfigure}
	\caption{Diagrammatic representation of (a) the Bethe-Salpeter equation 
for 
the electron-phonon vertex and (b) the phonon self-energy.}
	\label{fig:BS}
\end{figure}
Assuming the electron-phonon vertex to be independent of frequency and spin, 
the renormalized electron-phonon term is given by
\begin{align}
\mathcal{L}_{\rm f-ph} = \sum_{k, 
\mathbf q,\sigma=\uparrow,\downarrow}\Gamma(\mathbf q, \mathbf 
k)\psi^\dagger_{k + \mathbf q/2,\sigma}\psi_{k - 
\mathbf q/2,\sigma}\varphi(\mathbf{q}) + \mathrm{H.c.}
\end{align}
where the bare vertex $\Gamma_0(\mathbf{q},\mathbf{k}) = \lambda_\text{ph} 
g_{\rm ph}(\mathbf{q})$. For simplicity we expand the dependence on 
electron momenta $\mathbf k$ using basis functions,
\begin{equation}
\Gamma(\mathbf{q},\mathbf{k}) = \sum_{\ell = 0}^4 
\Gamma_\ell(\mathbf{q})f_\ell(\mathbf{k}),
\end{equation}
which are chosen keeping in mind the nearest-neighbor nature of the short range 
antiferromagnetic fluctuations and are given in Table~\ref{tab:basis}. The 
Bethe-Salpeter equation for the electron-phonon vertex is diagrammatically 
depicted in Fig.~\ref{fig:BS_a} and reads
\begin{align}
&\Gamma_j(\mathbf{q}) = \delta_{j0}\lambda_\text{ph} 
g_{\rm ph}(\mathbf{q}) - 
\sum_{\ell=0}^4\Gamma_\ell(\mathbf{q})M_{j\ell}(\mathbf{q}), \\
&M_{j\ell}(\mathbf{q}) = 
(1-\delta_{j0})\frac{3J}{2}\int\frac{d^2\mathbf{k}^\prime}{(2\pi)^2}\Pi(\mathbf{
k}^\prime,
\mathbf{q})f_j(\mathbf{k}^\prime)f_\ell(\mathbf{k}^\prime), \\
&\Pi(\mathbf{k}^\prime,\mathbf{q}) = 
\sum_{\alpha_1,\alpha_2=\pm}Z_{\mathbf{k}^\prime + \mathbf{q}/2}^{\alpha_1} 
Z_{\mathbf{k}^\prime - \mathbf{q}/2}^{\alpha_2} \frac{n_f(E_{\mathbf{k}^\prime 
+ \mathbf{q}/2}^{\alpha_1}) - 
n_f(E_{\mathbf{k}^\prime-\mathbf{q}/2}^{\alpha_2})}{E_{\mathbf{k}^\prime + 
\mathbf{q}/2}^{\alpha_1} - E_{\mathbf{k}^\prime - \mathbf{q}/2}^{\alpha_2}}. 
\end{align}
The antiferromagnetic interaction does not renormalize the $s$-wave component 
of the vertex. This can be achieved by adding density-density interactions. It 
is also possible to add further neighbor couplings. These should, however, not 
impact our conclusions in an essential way.

\begin{figure}
\centering
\begin{subfigure}[b]{3.1in}
		\includegraphics[width=3in]{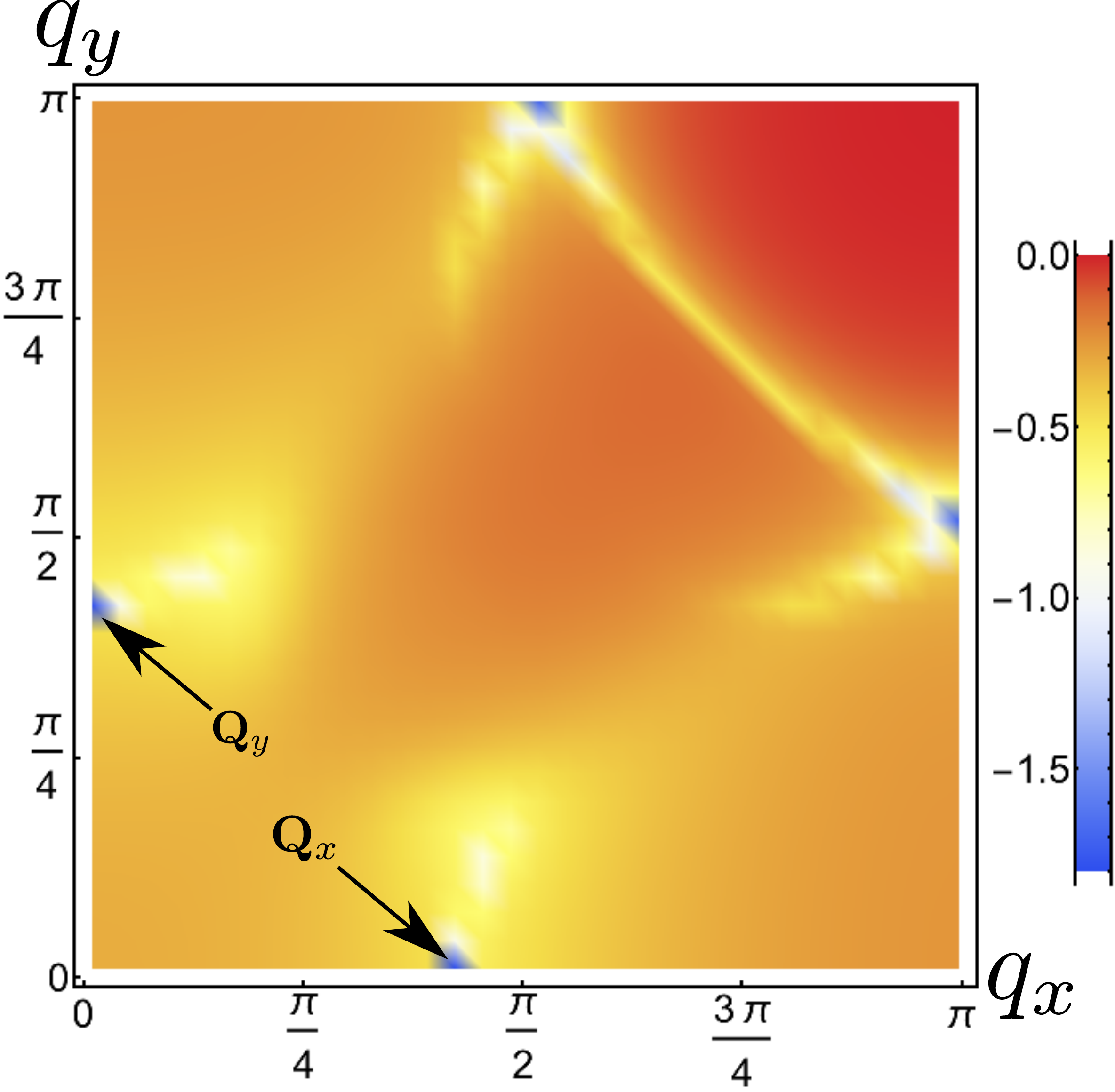}
		\caption{} \label{fig:PSEa}
	\end{subfigure}
	\vspace{0.25in}
\begin{subfigure}[b]{3.1in}
		\includegraphics[width=3in]{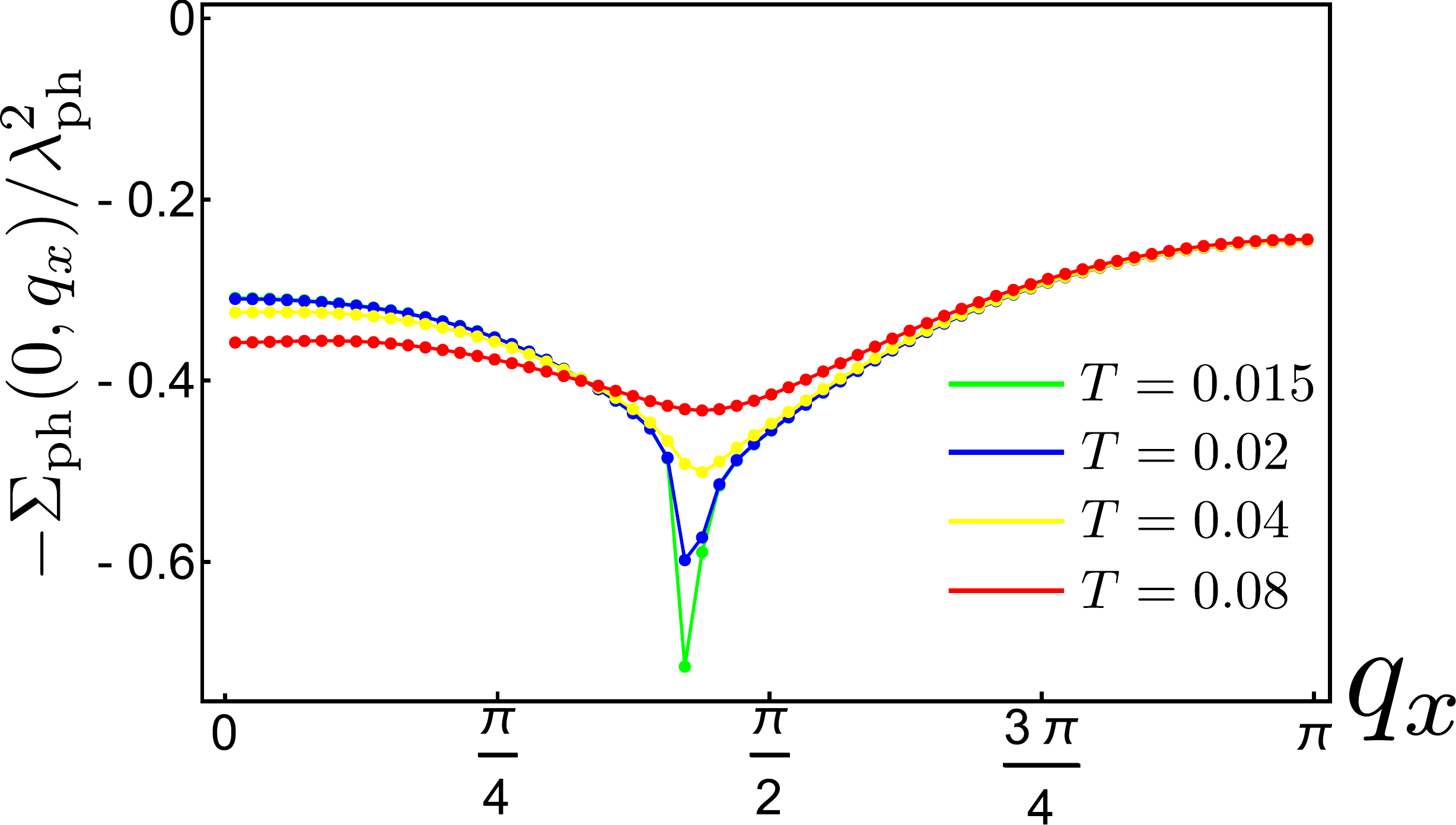}
		\caption{} \label{fig:PSEb}
	\end{subfigure}
\caption{(Color online) (a) Phonon self-energy $-\Sigma_{\rm 
ph}(0,\mathbf{q})/\lambda_{\rm ph}^2$ as a function of $\mathbf{q}$ for $J=2.07$
and $T = 0.01$.  The strongest softening occurs at the BDW wavevectors $\mathbf{Q}_x$ and 
$\mathbf{Q}_y$ which are marked in the figure. (b) Cuts along the $q_x$ axis of (a) for several 
temperatures. At $T=0.01$, the system is very close to the BDW instability for the value
of $J$ chosen.} 
\label{fig:PSE}
\end{figure}
The renormalized phonon dispersion $\omega_{\rm ph}(\mathbf q)$ is given by
\begin{equation}
	\omega_{\rm ph}^2(\mathbf{q}) 
= \omega_{\rm ph,0}^2 - \Sigma_{\rm ph}(0,\mathbf{q})
\end{equation}
where
\begin{equation}
\Sigma_{\rm ph}(0,\mathbf{q}) = -\lambda_\text{ph} 
g_{\rm ph}(\mathbf{q})\int \frac{d^2\mathbf{k}^\prime}{(2\pi)^2} 
\Gamma(\mathbf{q},\mathbf{k}^\prime)\Pi(\mathbf{k}^\prime,\mathbf{q})
\label{eq:PSE}
\end{equation}
is the static phonon self-energy computed with the renormalized electron-phonon 
vertex, as depicted in Fig.~\ref{fig:BS_b}. Close to a BDW instability, we 
obtain a significant softening of the phonon modes at wavevectors near 
$\mathbf{Q}_n$, as shown in Fig.~\ref{fig:PSE}. The softening gets more 
pronounced with decreasing temperature and/or increasing antiferromagnetic 
interaction $J$. A similar Bethe-Salpeter equation for the electron-BDW fluctuation 
vertex $\lambda_{\rm BDW}$ yields a strong enhancement of $\lambda_{\rm BDW}$
close to the BDW instability.

Once $T$ is low enough, the system becomes superconducting and the above 
analysis no longer applies. It thus cannot explain by itself the jump in phonon 
softening at $T_c$ noted in Ref.~\onlinecite{LeTacon2014}. However, it would be 
interesting to repeat their experiment in the presence of a magnetic field that 
suppresses superconductivity, in which case we would expect a stronger phonon 
softening with decreasing temperature and increasing BDW correlation length.

We now study the phonon dynamics described by Eq.~\eqref{eq:PhononAction}, 
where $\omega_{\rm ph, 0}$ is replaced by the renormalized phonon frequency 
$\omega_{\rm ph}(\mathbf{q})$. Since the dispersion minima now lie effectively 
at $\mathbf{Q}_n$, we write down the effective Lagrangian for these modes 
consistent with translation, inversion and $C_4$ rotation symmetry
\begin{align}
&\mathcal{L}_{\rm ph,eff} = \int d^2\mathbf{x} 
\Bigg[\sum_{n=\{x,y\}}\Bigg((\partial_t\varphi(\mathbf{Q}_n))^2-K^2|\nabla 
\varphi(\mathbf{Q}_n)|^2 \nonumber \\
&-(\omega_{\rm ph}(\mathbf{Q}_n)^2-rA_{\rm IR}\cos^2(\omega 
t))|\varphi(\mathbf{Q}_n)|^2-\frac{u}{4}|\varphi(\mathbf{Q}_n)|^4\Bigg) - 
u^\prime
|\varphi(\mathbf{Q}_x)|^2|\varphi(\mathbf{Q}_y)|^2\Bigg].
\end{align}
From this, we obtain the equations of motion of the fundamental modes 
\begin{align}
&\partial_t^2\varphi(\mathbf{Q}_x) + (\omega_{\rm ph}^2(\mathbf{Q}_x) - rA_{\rm 
IR}^2\cos^2(\omega t))\varphi(\mathbf{Q}_x) \nonumber \\
&+\frac{u}{2}\varphi(\mathbf{Q}_x)^2\varphi^\ast(\mathbf{Q}_x)+ 
u^\prime\varphi(\mathbf{Q}_x)|\varphi(\mathbf{Q}_y)|^2 + 
\gamma(\mathbf{Q}_x)\partial_t
\varphi(\mathbf{Q}_x)= 0,
\end{align}
and the partner equations for $x\leftrightarrow y$ and 
$\varphi\leftrightarrow\varphi^\ast$, where we also added Landau damping terms 
representing effects of the imaginary part of the phonon self-energy, which we 
have not computed explicitly. This Landau damping will be enhanced by the large 
value of $\Gamma(\mathbf{Q}_n,\mathbf{k})$ close to the BDW instability, leading 
to spectral broadening at those wavevectors, as seen in experiments. For 
simplicity in illustrating qualitative features, we consider $C_4$ symmetric 
initial conditions with real mode values and $u^\prime=u/2$. The equations then 
read
\begin{align}
&\partial_t^2\varphi(\mathbf{Q}_n) + (\omega_{\rm ph}^2(\mathbf{Q}_n) - rA_{\rm 
IR}^2\cos^2(\omega t))\varphi(\mathbf{Q}_n) + u\varphi(\mathbf{Q}_n)^3 + 
\gamma(\mathbf{Q}_n)\partial_t
\varphi(\mathbf{Q}_n)= 0.
\label{eq:Mathieu}
\end{align}

\begin{figure}
\begin{subfigure}[c]{3.1in}
		\includegraphics[width=3in]{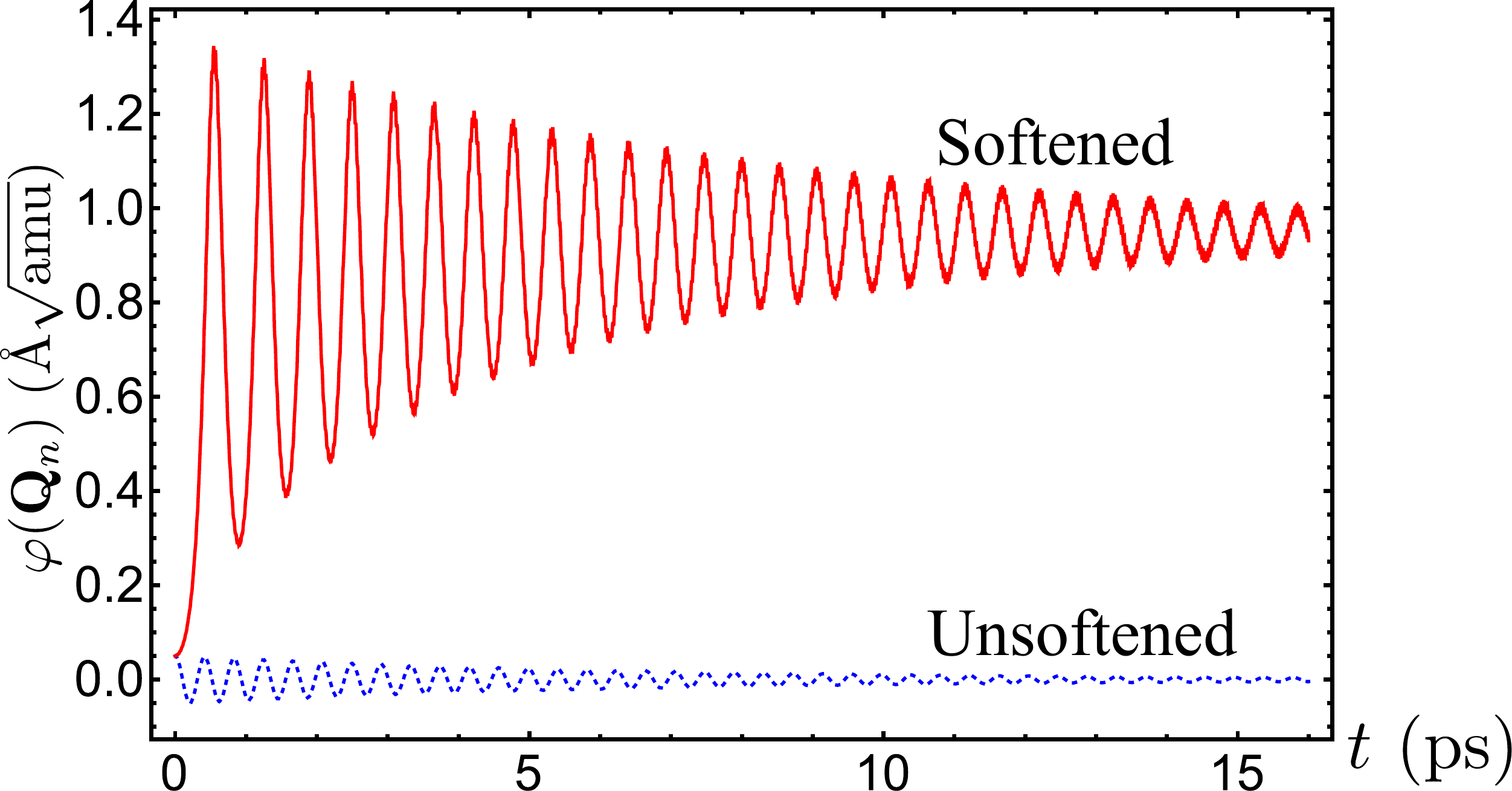}
		\caption{} \label{fig:oscillationsa}
	\end{subfigure}
	\vspace{0.25in}
\begin{subfigure}[c]{3.1in}
		\includegraphics[width=3.3in]{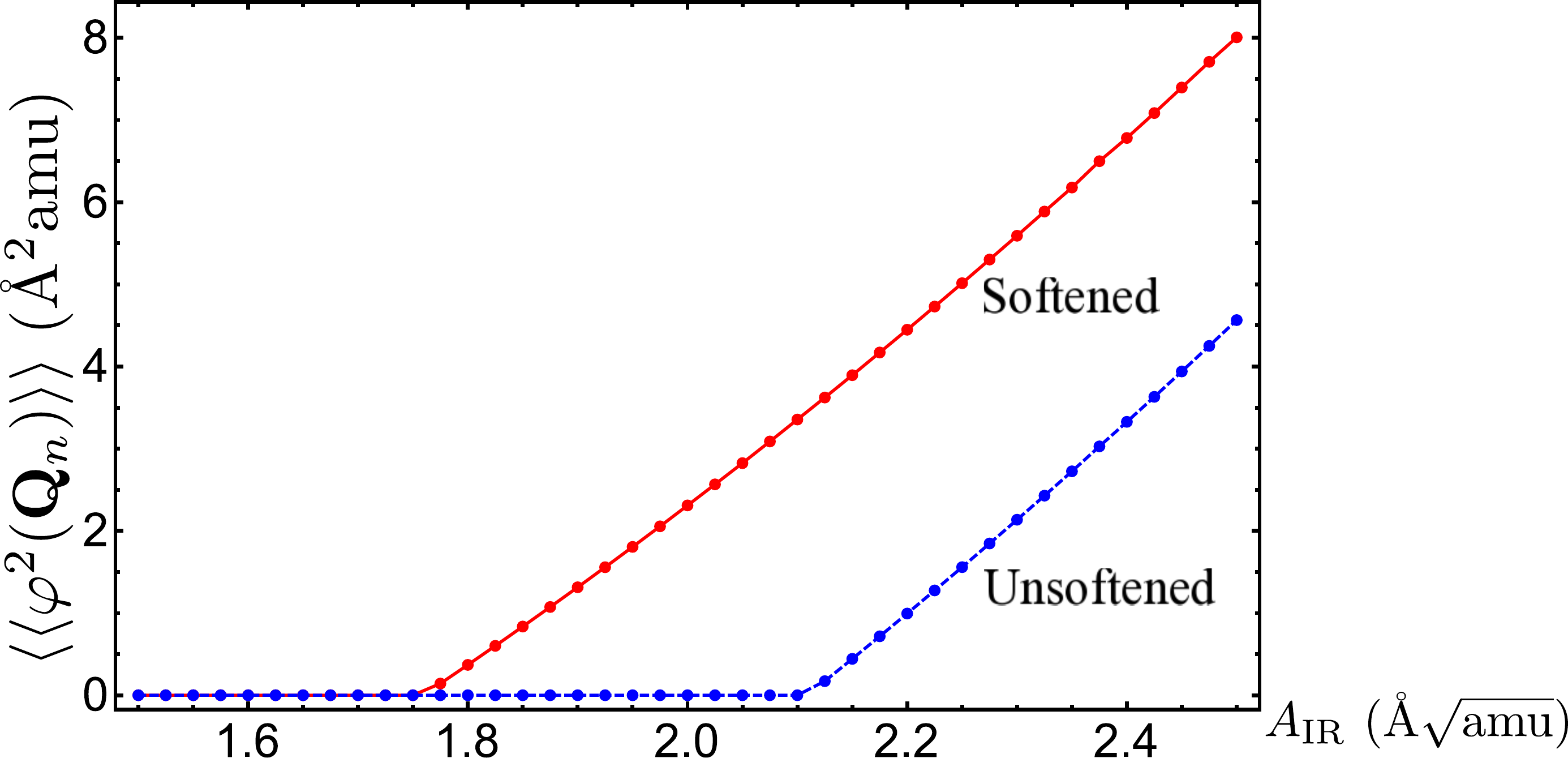}
		\caption{} \label{fig:oscillationsb}
	\end{subfigure}
\caption{(Color online) (a) Phonon dynamics for unsoftened phonons with 
$\nu<\nu_c$ (dashed, blue) and for softened phonons with $\nu>\nu_c$ (solid, 
red). We use parameters from Ref.~\onlinecite{Subedi2014}, a 
softening of $\sim$15\% and a small damping: The phonon frequencies are 
$\omega_{\rm ph}(\mathbf{Q}_n)^2 
=103.55~\mathrm{meV/\mbox{\AA}^2/amu}=\omega_{\rm ph,0}^2$ for the unsoftened 
phonon (dashed, blue) and
$\omega_{\rm ph}^2(\mathbf{Q}_n) = 72.55~\mathrm{meV/\mbox{\AA}^2/amu}$ for 
the softened phonon (solid, red) and other parameters are $\omega^2 = 
1462.3~\mathrm{meV/\mbox{\AA}^2/amu}$, $r = 
46.98~\mathrm{meV/\mbox{\AA}^4/amu^2}$, $u =  
8.36~\mathrm{meV/\mbox{\AA}^4/amu^2}$, $\gamma^2(\mathbf{Q}_n) = 
0.01~\mathrm{meV/\mbox{\AA}^2/amu}$ and $A_{\rm IR} = 
1.85~\mathrm{\mbox{\AA}\sqrt{amu}}$. 
The value of the softening used corresponds to an energy difference of about 3 
meV. (b) The mean-square amplitude of the phonons as a function of the driving 
amplitude $A_{\rm IR}$.}
\label{fig:oscillations}
\end{figure}

These equations exhibit the well known Mathieu 
instability~\cite{AbramowitzStegun, Subedi2014, Jordan1999} for large enough 
driving amplitudes, when the control parameter $\nu$ exceeds a critical value,
\begin{equation}
\nu = \frac{r A_{\rm IR}^2}{2\omega_{\rm ph}^2(\mathbf{Q}_n)} > 
\nu_c = 1 + \frac{\omega_{\rm ph}^2(\mathbf{Q}_n)}{8\omega^2} + ...~.
\end{equation}
The instability happens first for the softened phonons for $r > 0$. The 
anharmonic coupling $u > 0$ stabilizes the system and allows the phonon 
potential to acquire a double well shape. The phonons then oscillate about the 
minima of the double well instead of about zero, and the damping terms simply 
damp out these oscillations (see Fig.~\ref{fig:oscillationsa}). In this 
regime, an average displacement of the Raman mode is possible. 
Equation~\eqref{eq:Mathieu} also shows an instability for $r < 
0$~\cite{Jordan1999}, but in this case it does not happen first for the softest 
phonons and the physics is different.

Figure~\ref{fig:oscillations} shows examples for the phonon dynamics in 
different regimes, highlighting the impact of phonon softening. The parameters 
used in this figure are taken from Ref.~\onlinecite{Subedi2014}. Since $\nu$ 
increases when $\omega_{\rm ph}(\mathbf{Q}_n)$ is softened, it is possible that 
a significant softening for Raman phonons close to the BDW instability can lead 
to the modes near $\mathbf{Q}_n$ being in the $\nu>\nu_c$ regime, thereby 
acquiring large oscillation amplitudes. Equivalently, if $A_{\rm IR}$ is large 
enough, these modes will enter this regime. Thus, phonon anharmonicities in 
combination with momentum-specific phonon softening can lead to a strong driving 
of Raman modes close to the BDW ordering wavevectors. The dynamics of phonon 
modes at momenta away from $\mathbf{Q}_n$, such as those near $\mathbf{q}=0$, 
is qualitatively similar to that of the unsoftened mode in 
Fig.~\ref{fig:oscillations}. In Fig.~\ref{fig:oscillationsb}, we show the 
time-averaged Raman displacement after long excitation times. The maximum of 
the time-averaged Raman displacement after excitation pulses behaves in a 
qualitatively similar way. It is interesting to note that a similar 
threshold behavior is also observed in experiments~\cite{Kaiser2014}.

\section{Coupling of phonons to bond-density wave and pairing fluctuations}
\label{sec:box}

We now investigate the effect of the proposed strongly driven Raman modes near 
$\mathbf{Q}_n$ on BDW and dSC fluctuations in the pseudogap phase. To do this, 
we integrate out the fermions, producing corrections to the effective action of 
the phenomenological model describing the competition of these order 
parameters. 
\begin{figure}
	\begin{subfigure}[c]{0.3\linewidth}
		\includegraphics{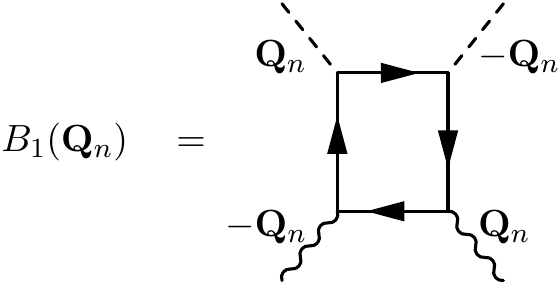}
		\caption{} \label{fig:boxd_a}
	\end{subfigure}
\hspace{1cm}
	\begin{subfigure}[c]{0.3\linewidth}
		\includegraphics{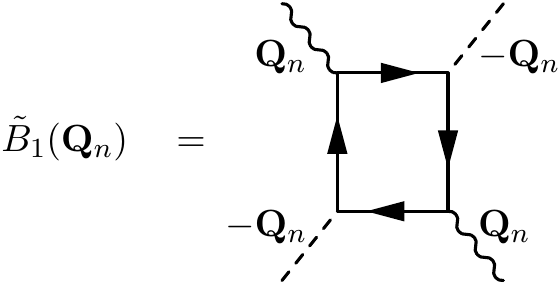}
		\caption{} \label{fig:boxd_b}
	\end{subfigure}

\begin{subfigure}[c]{0.3\linewidth}
		\includegraphics{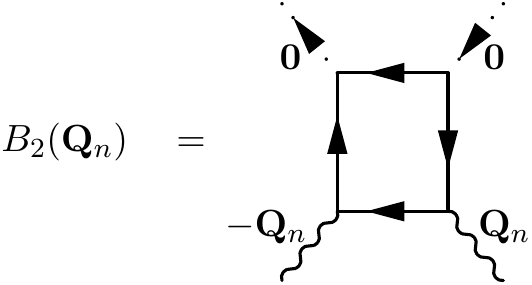}
		\caption{} \label{fig:boxd_c}
	\end{subfigure}
\hspace{1cm}
	\begin{subfigure}[c]{0.3\linewidth}
		\includegraphics{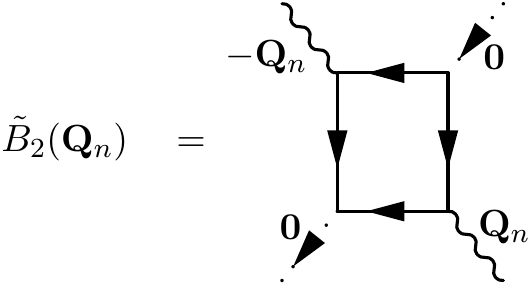}
		\caption{} \label{fig:boxd_d}
	\end{subfigure}
	\caption{Box diagrams yielding the lowest-order couplings between phonons 
and BDW as well as pairing fluctuations. Wavy lines represent phonons, dashed 
and dotted lines BDW and dSC fluctuations, respectively.}
	\label{fig:boxd}
\end{figure}
 The simplest corrections are given by the ``box'' 
diagrams~\cite{Chowdhury2014} shown in Fig.~\ref{fig:boxd}. We consider only 
diagrams where the fermions are close to the Fermi surface throughout all stages 
of the scattering process, as these are expected to be most relevant at low 
temperatures. The resulting mass renormalization for BDW and dSC fluctuations 
reads
\begin{align}
\delta S_{\rm cl} = T\int d^2\mathbf{x} 
\sum_{n=x,y}\langle\langle\varphi(\mathbf{Q}_n)^2\rangle\rangle&\Big[
B_1(\mathbf{Q}_n) |\Phi_n|^2 + \tilde{B}_1(\mathbf{Q}_n) 
\operatorname{Re}[\Phi_n^2] \nn
&+ \left(B_2(\mathbf{Q}_n) + 
\tilde{B}_2(\mathbf{Q}_n)\right)|\Psi|^2\Big],
\end{align}
where the $\langle\langle~\rangle\rangle$ denotes time-averaging over an 
oscillation cycle. $\langle\langle\varphi(\mathbf{Q}_n)^2\rangle\rangle$ 
depends on $A_{\rm IR}$ as shown in Fig.~\ref{fig:oscillationsb}. For 
simplicity we ignore specifics of the transient behavior in this work. We are 
also making the assumption that only the low-momentum modes of the NLSM are 
important, and their mass renormalization is given by that of the zero modes. 
The box integrals are then given by
\begin{align}
&B_1(\mathbf{Q}_n)=\tilde{B}_1(\mathbf{Q}_n)= \nn
&\lambda_{\rm BDW}^2 \lambda_{\rm ph}^2\int\frac{d^2\mathbf{k}}{(2\pi)^2} 
T\sum_{\omega_n}\sum_{\alpha_{1-4}=\pm}\frac{Z_\mathbf{k}^{\alpha_1}Z_{\mathbf{k
} + \mathbf{Q}_n}
^{\alpha_2}Z_\mathbf{k}^{\alpha_3}Z_{\mathbf{k} + 
\mathbf{Q}_n}^{\alpha_4}g_d(\mathbf{k} + 
\mathbf{Q}_n/2)^2 g_\text{ph}(\mathbf{Q}_n)^2}{(i\omega_n - 
E_\mathbf{k}^{\alpha_1})(i\omega_n - 
E_{\mathbf{k} + \mathbf{Q}_n}^{\alpha_2})(i\omega_n - 
E_\mathbf{k}^{\alpha_3})(i\omega_n-E_{\mathbf{k} + \mathbf{Q}_n}^{\alpha_4})}, 
\end{align}

\begin{align}
&B_2(\mathbf{Q}_n)=-\lambda_{\rm dSC}^2 \lambda_{\rm ph}^2\times\nonumber \\
&\times \int\frac{d^2\mathbf{k}}{(2\pi)^2}T\sum_{\omega_n} 
\sum_{\alpha_{1-4}=\pm}\frac{Z_\mathbf{k}^{\alpha_1}Z_{\mathbf{k} + 
\mathbf{Q}_n}^{\alpha_2}Z_{\mathbf{k} + 
\mathbf{Q}_n}^{\alpha_3}Z_{\mathbf{k} + \mathbf{Q}_n}^{\alpha_4}g_d(\mathbf{k} 
+ \mathbf{Q}_n)^2 g_\text{ph}(\mathbf{Q}_n)^2}{(i\omega_n-E_\mathbf{k}
^{\alpha_1})(i\omega_n - E_{\mathbf{k} + \mathbf{Q}_n}^{\alpha_2})(i\omega_n + 
E_{\mathbf{k} + \mathbf{Q}_n}^{\alpha_3})(i\omega_n - E_{\mathbf{k} + 
\mathbf{Q}_n}^{\alpha_4})}, \\
&\tilde{B}_2(\mathbf{Q}_n)=\lambda_{\rm dSC}^2 \lambda_{\rm ph}^2 
\int\frac{d^2\mathbf{k}}{(2\pi)^2}T\sum_{\omega_n} 
\sum_{\alpha_{1-4}=\pm}\frac{Z_\mathbf{k}^{\alpha_1} 
Z_{\mathbf{k}+\mathbf{Q}_n}^{\alpha_2}
Z_{\mathbf{k} + \mathbf{Q}_n}^{\alpha_3}Z_\mathbf{k}^{\alpha_4}g_d(\mathbf{k})
g_d(\mathbf{k} + \mathbf{Q}_n) g_\text{ph}(\mathbf{Q}_n)^2}{(i\omega_n - 
E_\mathbf{k} ^{\alpha_1})(i\omega_n - E_{\mathbf{k} + 
\mathbf{Q}_n}^{\alpha_2})(i\omega_n + E_{\mathbf{k} + 
\mathbf{Q}_n}^{\alpha_3})(i\omega_n + E_{\mathbf{k}}^{\alpha_4})},
\end{align}
where we have exploited the evenness of certain quantities to simplify the 
expressions. Note that if the contribution to these diagrams comes mostly
from ``hot" regions of the Fermi surface that are nested by $\mathbf{Q}_n$,
where $E^\pm_{\mathbf{k}+\mathbf{Q}_n}\approx-E^\pm_{\mathbf{k}}$,
$Z^\pm_{\mathbf{k}+\mathbf{Q}_n}\approx Z^\pm_{\mathbf{k}}$, and
$g_d(\mathbf{k}+\mathbf{Q}_n)\approx g_d(\mathbf{k})$, 
$\mathbf{Q}_n$, $B_2$ and $\tilde{B}_2$ will nearly cancel.

\begin{figure}
\begin{subfigure}[c]{3.1in}
		\includegraphics[width=3in]{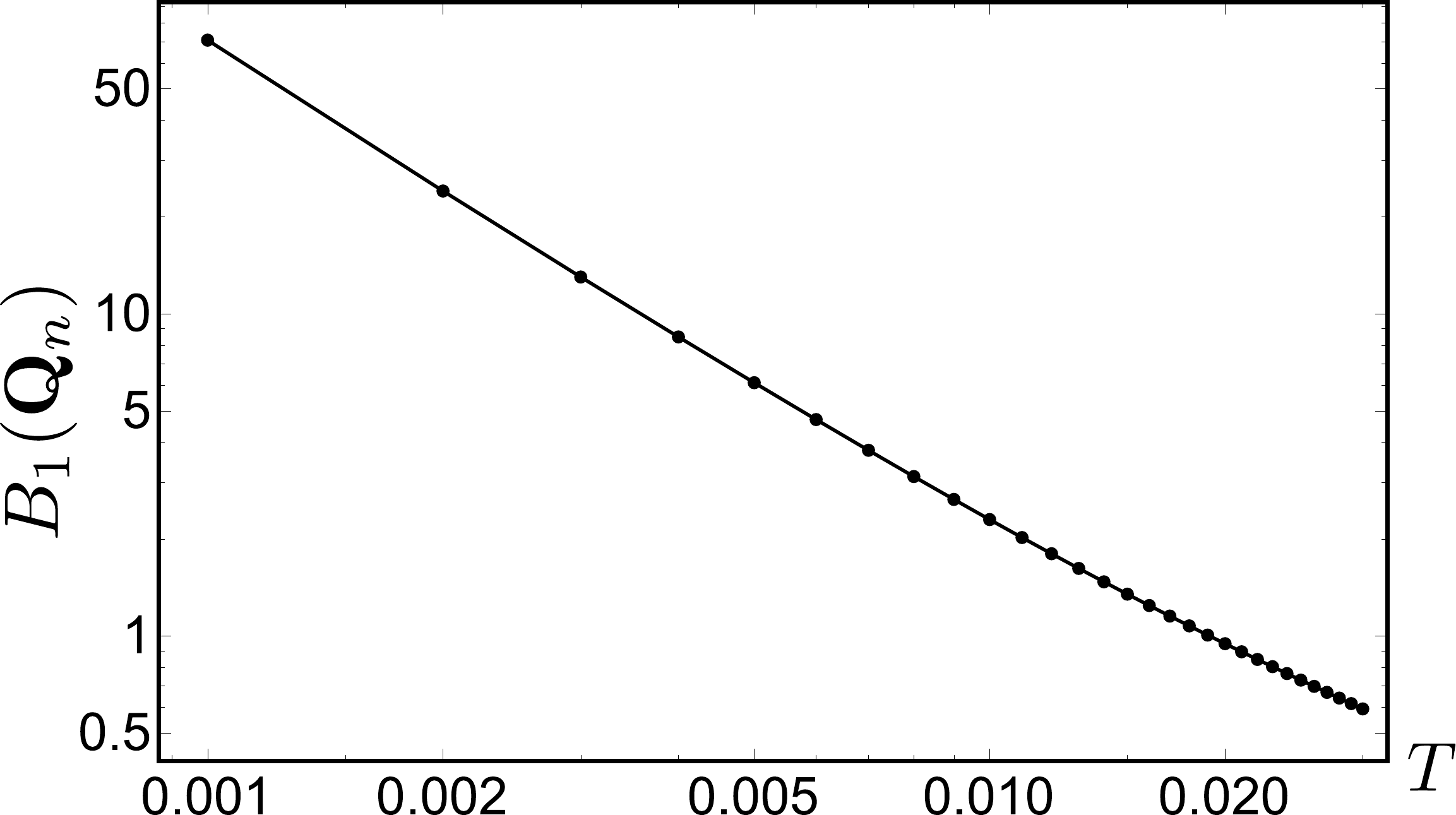}
		\caption{} \label{fig:boxTa}
	\end{subfigure}
	\vspace{0.25in}
\begin{subfigure}[c]{3.1in}
		\includegraphics[width=3in]{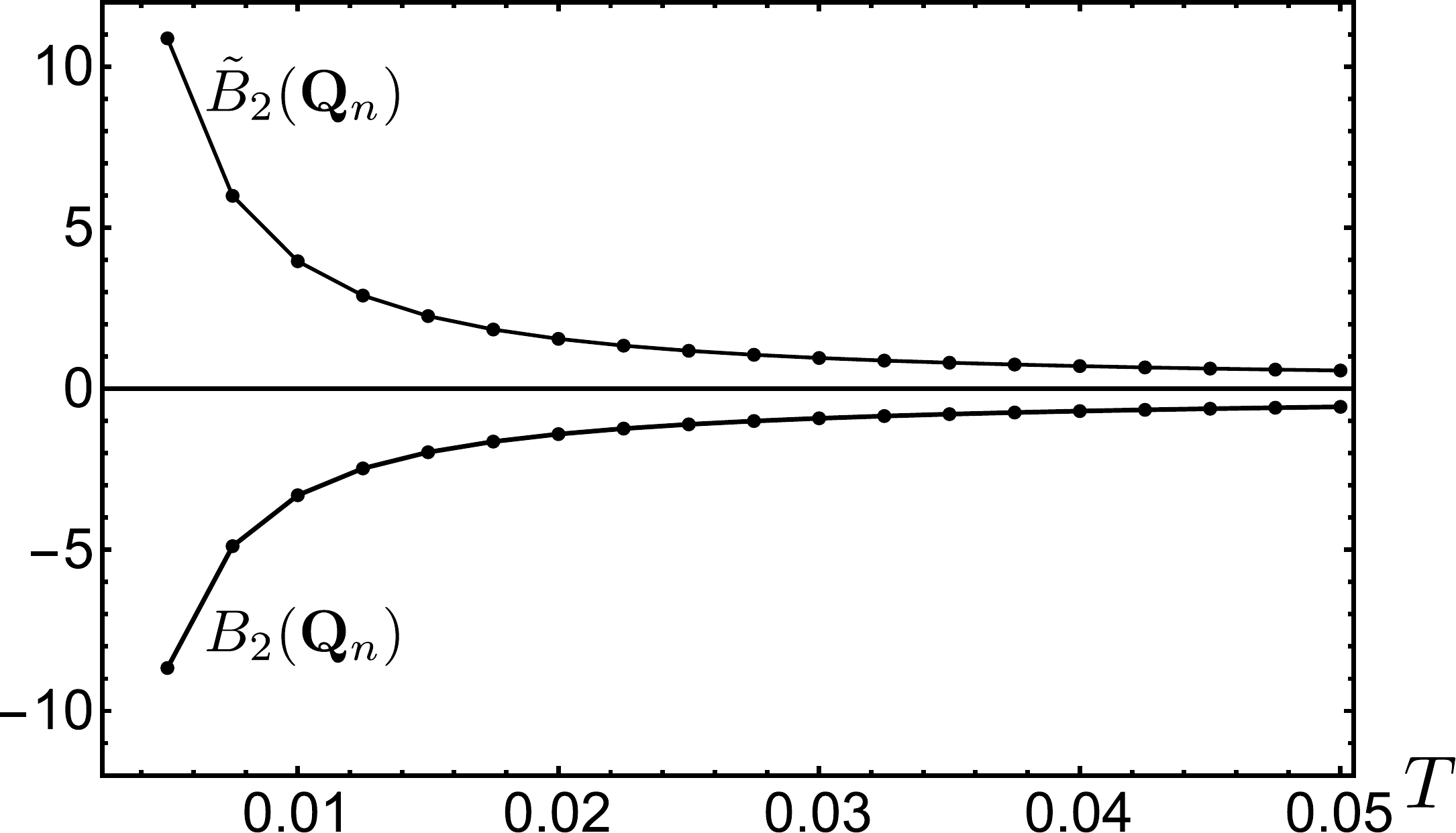}
		\caption{} \label{fig:boxTb}
	\end{subfigure}

\begin{subfigure}[c]{3.1in}
		\includegraphics[width=3in]{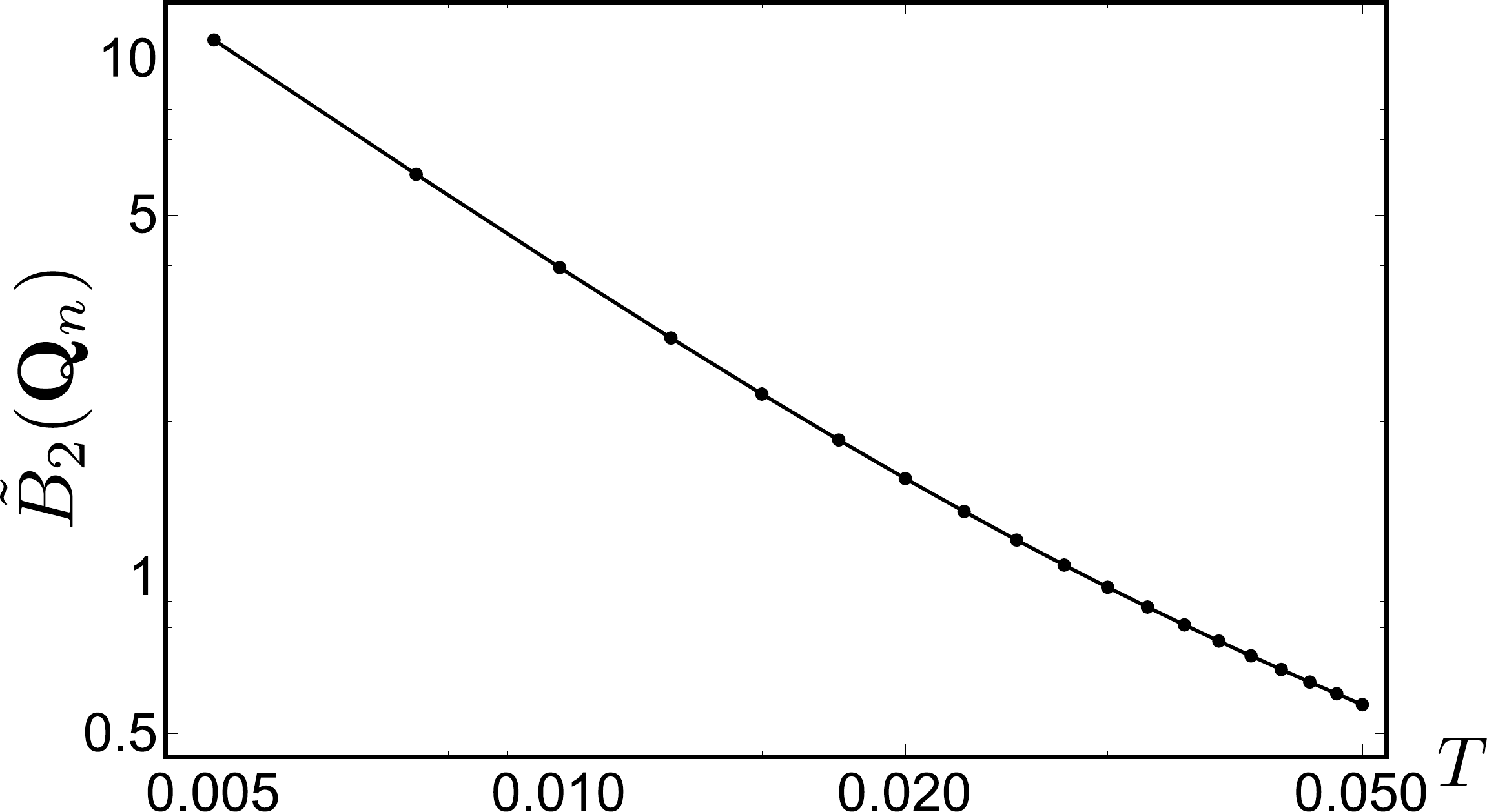}
		\caption{} \label{fig:boxTc}
	\end{subfigure}
\caption{Temperature dependence of box diagrams: (a) $B_1(\mathbf{Q}_n)$, (b) 
$B_2(\mathbf{Q}_n)$ and $\tilde{B}_2(\mathbf{Q}_n)$, (c) 
$\tilde{B}_2(\mathbf{Q}_n)$. Note that $B_2(\mathbf{Q}_n) \approx 
-\tilde{B}_2(\mathbf{Q}_n)$. We used momentum-independent electron-phonon
vertices for simplicity and set $\lambda_{\rm BDW} = \lambda_{\rm dSC} 
= \lambda_{\rm ph} = 1$ for the purpose of visualization.}
\label{fig:boxT}
\end{figure}

Let us start by deriving the low-temperature scaling of these diagrams. At 
very low temperatures, we can set $\alpha_{1-4}=-$. The corrections to the box 
integrals from the terms excluded by doing this are at most $O(T/\Delta)$, 
where $\Delta\sim~E_\mathbf{k}^+-E_\mathbf{k}^-\gg T$. At low temperatures, the 
integrands of the box diagrams are strongly peaked at the hot regions near the 
tips of the hole pockets. By making a gradient expansion of the dispersion at the pocket tips 
connected by for example $\mathbf Q_y$, which yields $E_\mathbf{k}^- \sim -v 
k_y + \kappa k_x^2$ and $E_{\mathbf{k}+\mathbf{Q}_y}^-\sim v k_y + \kappa 
k_x^2$ (see Fig.~\ref{fig:FS}), we can estimate the low-temperature behavior of 
the diagrams. The form factors and quasiparticle residues are slowly varying 
near the tips and may thus be approximated by the value at the tips of the 
pockets. The box integrals still converge in this limit, and a simple rescaling 
of momenta, $k_y\rightarrow T k_y$ and $k_x \rightarrow \sqrt{T}k_x$, then 
yields $B_1,~B_2,~\tilde{B}_2\sim T^{-3/2}$. The corrections to the mass terms 
in the effective action hence scale as $T^{-1/2}$. Also, since particle-hole 
symmetry holds to linear order in $\mathbf{k}$, there is a near cancellation 
between $B_2$ and $\tilde{B}_2$. These statements are confirmed by numerical 
evaluation of the diagrams without these restrictions, as shown in 
Fig.~\ref{fig:boxT}. The temperature dependence of the renormalized 
vertices or the corrections with $d$-wave form factors do not significantly 
affect the scaling and the cancellation between $B_2$ and $\tilde{B}_2$. We 
find $B_1(\mathbf{Q}_n) > 0$ and $B_2(\mathbf{Q}_n) \approx - 
\tilde{B}_2(\mathbf{Q}_n) < 0$.

The leading corrections to the effective action thus simplify to
\begin{align}
&\delta S_{\rm cl} \approx T\int d^2\mathbf{x} 
\sum_{n=x,y}\langle\langle\varphi(\mathbf{Q}_n)^2\rangle\rangle 
B_1(\mathbf{Q}_n) \operatorname{Re}[\Phi_n]^2,
\end{align}
yielding a mass enhancement for BDW fluctuations along the 
$\operatorname{Re}[\Phi_n]$ directions (\ie\ $m^2 \rightarrow m^{\prime 2} = 
m^2 + \frac{2T}{\rho_s}\langle\langle\varphi(\mathbf{Q}_n)^2\rangle\rangle 
B_1(\mathbf{Q}_n)$). In order to demonstrate that this is indeed favorable for 
superconductivity, we assume that the time-averaged behavior is equivalent to an 
effective equilibrium model for which we can compute the enhancement of $T_c$. 
This is certainly only indicative for the non-equilibrium situation, and should 
not be taken as a serious quantitative estimate of the strength of any 
transient superconducting state. True transient superconducting behavior can 
only be understood by solving the full time-dependent problem and studying the 
behavior of the superfluid density, which is beyond the scope of this work. Note 
that within our approximations, the coupling of order parameters to driven 
phonons acts as a parameter quench for the non-linear sigma model. The results 
of Fu~\etal~\cite{WFu2014} then suggest that the qualitative trends in the 
time-averaged results carry over to a solution of the equations of motion. The 
$T_c$ enhancement is computed by condensing one component of the dSC order 
parameter and neglecting fluctuations that would vanish in a large-$N$ limit, as 
done in Ref.~\onlinecite{SubirsBook}. The gap equation reads 
\begin{equation}
\Psi_0^2 = 1 - 
\frac{T}{3\rho_s}\left(\int^\Lambda\frac{d^2\mathbf{k}}{(2\pi)^2}\frac{1}{k^2} 
+ \int^\Lambda\frac{d^2\mathbf{k}}{(2\pi)^2}\frac{1}{\eta k^2 + m^2} + 
\int^\Lambda\frac{d^2\mathbf{k}}{(2\pi)^2}\frac{1}{\eta k^2 + m^{\prime 
2}}\right) = 1- \frac{T}{T_c}.
\end{equation}
This can be regulated and the $T_c$ enhancement computed by differentiating and 
re-integrating with respect to $m^{\prime 2}$, yielding
\begin{equation}
T_c = \frac{T_c^{(0)}}{1 - 
\frac{T_c^{(0)}}{12\pi\rho_s\eta}\ln\left(\frac{m^{\prime 2}}{m^2}\right)}.
\end{equation}
The logarithm suggests that the enhancement of superconductivity is a modest 
effect. Nevertheless, it is a correction in the correct direction, which is 
basically what we wanted to demonstrate. 

In the appendix, we present further results for the momentum dependence of the 
box diagrams (Appendix~\ref{sec:Appendix:WaveVectors}) and for the influence of 
the geometry of the Fermi surface (Appendix~\ref{sec:Appendix:LargeFS}). The 
sign structure of the box diagrams implies that it would be detrimental for the 
proposed mechanism of enhancing superconductivity if phonons with other momenta 
than the BDW wave vector are strongly driven. Moreover, BDW with axial wave 
vectors cannot be efficiently melted in a system with a large Fermi surface. 
However, for BDW with diagonal wave vectors conclusions similar to those in 
this section hold.

\section{Discussion and Conclusions}
\label{sec:disc}
In summary, we proposed a theory for light-induced superconductivity and phonon 
renormalization in slightly underdoped cuprates below the onset temperature for 
BDW fluctuations. We described the competition between superconductivity and 
BDW fluctuations by a phenomenological model and studied how the interplay 
between the two orders is influenced by externally driven phonons. At an 
incipient BDW instability in a system with a small Fermi surface, as in a 
fractionalized Fermi liquid, the electron-phonon vertex is strongly enhanced, 
which leads to a significant softening of the phonon dispersion at the BDW wave 
vector. It is expected that the softening due to BDW fluctuations is also 
accompanied by a strongly enhanced phonon linewidth, the latter being observed 
in experiment~\cite{LeTacon2014}.

The phonon softening at the BDW wave vector entails large phonon oscillations or 
displacements when the system is sufficiently strongly driven with light. Due to 
the strong coupling between phonons at the BDW wave vector and BDW fluctuations, 
this leads to an efficient melting of BDW correlations by renormalizing the mass 
of BDW fluctuations. The coupling between phonons and Cooper pairs is 
significantly smaller because certain quartic couplings between them nearly 
cancel. This asymmetry regarding the mass renormalization for collective 
excitations tips the balance between competing orders towards pairing 
correlations when the system is stimulated with light. A time-averaged 
approximation for the phenomenological model of competing orders yields an 
enhancement of $T_c$, which may give rise to a transient superconducting state 
in a dynamical treatment. The latter is, however, beyond the scope of the 
present work.

The proposed mechanism is applicable only below the onset temperature of 
BDW fluctuations and cannot explain the transient superconducting state that is 
found up to room temperature. Our theory for phonon renormalizations could be 
tested by repeating the experiments by Le~Tacon~\etal~\cite{LeTacon2014} in a 
magnetic field slightly below the zero-field equilibrium $T_c$. Suppressing 
superconductivity should make BDW fluctuations more critical and yield a 
stronger renormalization of the phonon dispersion and linewidth, leading to a 
continuation of the trends that are found above $T_c$. Another test of our 
theory would be possible by measuring the momentum-space power spectrum of the 
driven phonons in the transient superconducting state. According to our theory, 
the (Raman) phonons at the BDW wavevector should be excited much more strongly 
than those at other wavevectors. The proposed mechanism for enhanced 
superconductivity could presumably work collaboratively with the mechanism 
involving interlayer hopping suggested by Raines~\etal~\cite{Raines2015}, which 
would enhance the effect.

This work also constitutes a first step towards an understanding of controlling 
competing order parameters by driving phonons with light. A theory similar to 
the one suggested in this work could be applicable to stripe ordered 
Lanthanum-based cuprates~\cite{Hunt2015}, in which strong phonon softening is 
also observed~\cite{Reznik2006}. The similarities regarding two-dimensional 
incommensurate charge correlations, anomalous strength of phonon softening and 
role of anharmonicity effects in YBCO and $\mathrm{NbSe}_2$~\cite{Leroux2015} 
provoke the question whether it would also be possible to melt CDW order by 
excitation of phonons with light in the latter material.

\begin{acknowledgments}
We acknowledge valuable discussions with D.~Chowdhury, 
S.~Kaiser and S.~Sachdev. This research was supported by the NSF under Grant 
DMR-1360789 and the German National Academy of Sciences Leopoldina through grant 
LPDS~2014-13.
\end{acknowledgments}

\appendix

\section{Other Phonon Wavevectors}
\label{sec:Appendix:WaveVectors}

\begin{figure}
\begin{subfigure}[c]{3.1in}
		\includegraphics[width=3in]{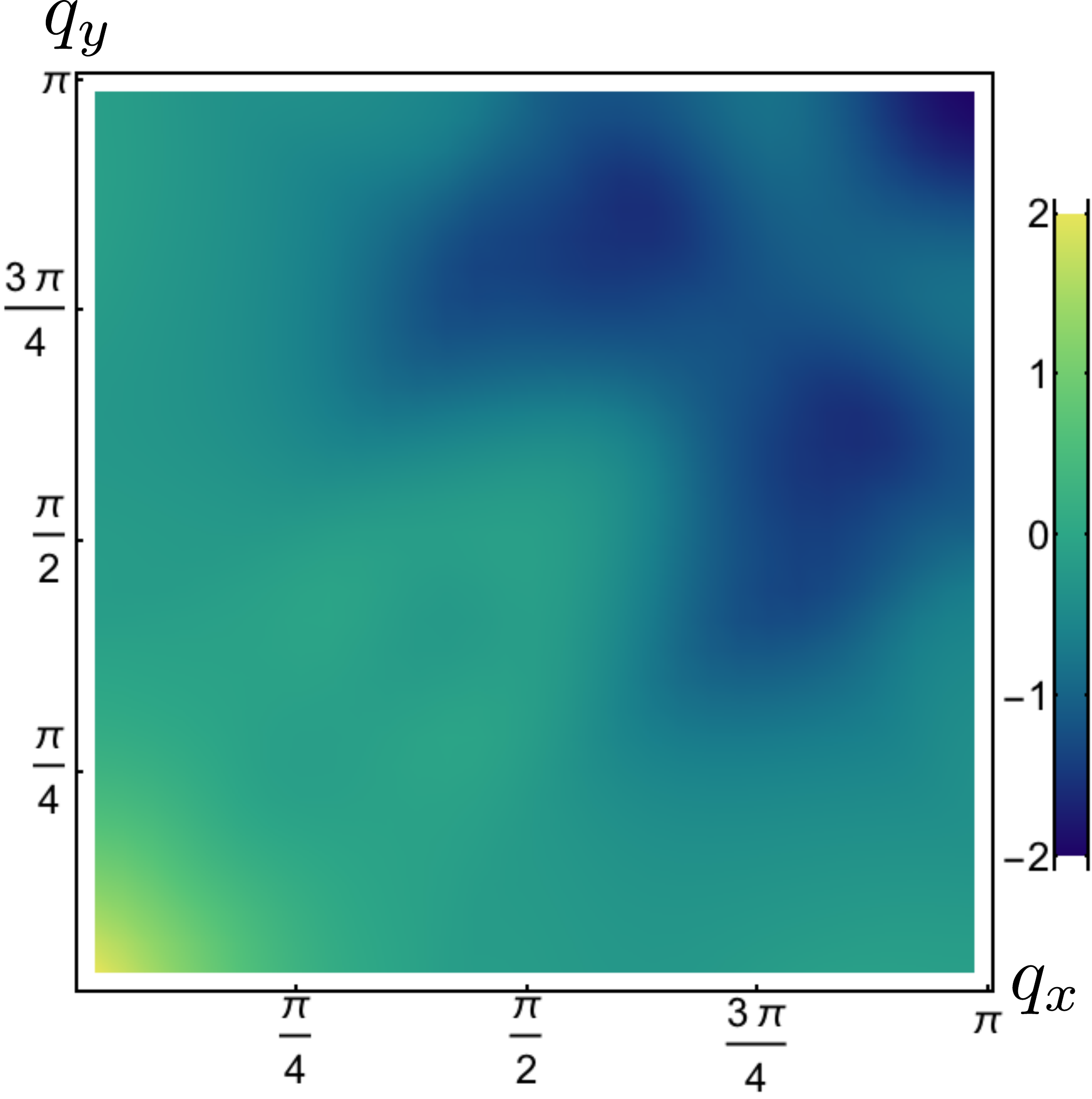}
		\caption{} \label{fig:boxqa}
	\end{subfigure}
	\vspace{0.25in}
\begin{subfigure}[c]{3.1in}
		\includegraphics[width=3in]{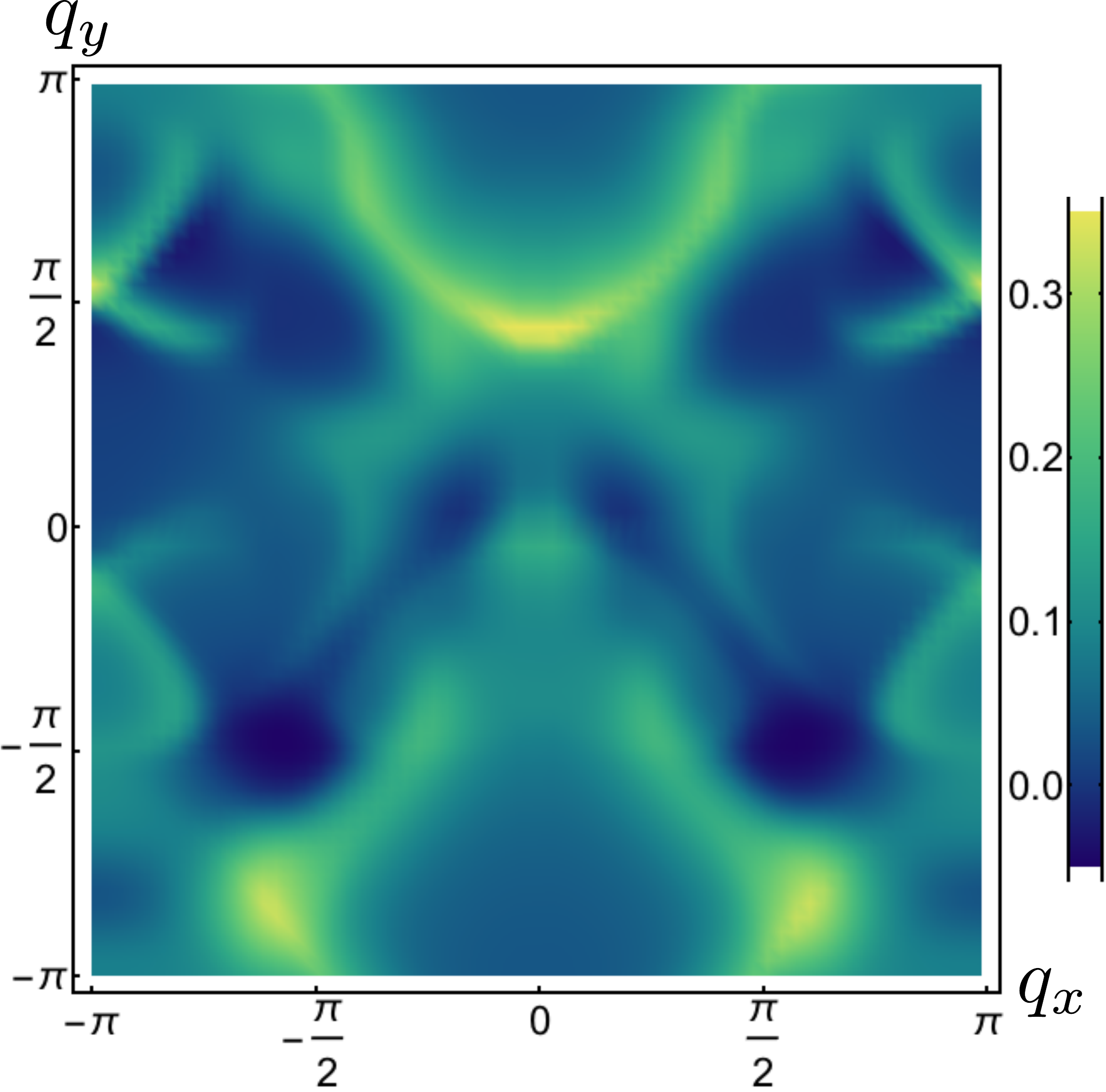}
		\caption{} \label{fig:boxqb}
	\end{subfigure}
\caption{(Color online) Momentum dependence of box diagrams: (a) 
$B_2(\mathbf{q}) + \tilde{B}_2(\mathbf{q})$ and (b) $B_1(\mathbf{q})$. The 
temperature is $T=0.05$ in both plots. We have again used momentum-independent
electron-phonon vertices and set $\lambda_{\rm BDW} = \lambda_{\rm dSC} = \lambda_{\rm ph} = 1$ for the purpose of visualization.}
\label{fig:boxq}
\end{figure}
In this appendix we present results for the momentum dependence of quartic 
couplings of order parameters and certain other phonon modes, although we 
deduced that these are not as strongly driven as the modes near $\mathbf{Q}_n$. 
For simplicity we approximate the electron-phonon vertex by a momentum-independent one 
(we are only interested in the qualitative behavior of the box integrals across the 
Brillouin zone). Then the dependence on the phonon momentum $\mathbf q$ of the box 
diagram contributing to the dSC mass renormalization ($B_2(\mathbf{q}) + 
\tilde{B}_2(\mathbf{q})$) is shown in Fig.~\ref{fig:boxqa}. It is close to zero 
or even negative, except when $\mathbf{q}$ is near zero, where it is strongly 
positive due to nesting all over the Fermi surface. Hence, it is essential that 
the phonon modes near $\mathbf{q}=0$ are not strongly driven for the proposed 
mechanism for light-induced superconductivity to work. This can be tested in an 
experiment by measuring the momentum-space power spectrum of the driven phonons.

\begin{figure}
	\begin{subfigure}[c]{0.25\linewidth}
		\includegraphics{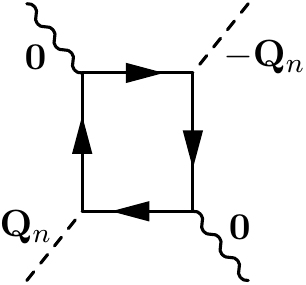}
		\caption{} \label{fig:boxmisc_a}
	\end{subfigure}
\hspace{1cm}
	\begin{subfigure}[c]{0.25\linewidth}
		\includegraphics{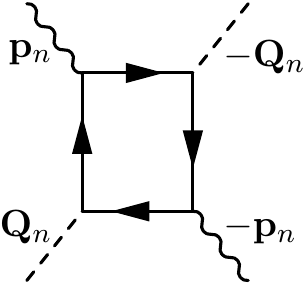}
		\caption{} \label{fig:boxmisc_b}
	\end{subfigure}
\hspace{1cm}
\begin{subfigure}[c]{0.25\linewidth}
		\includegraphics{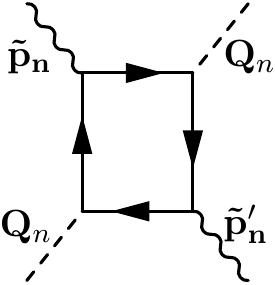}
		\caption{} \label{fig:boxmisc_c}
	\end{subfigure}
	\caption{Some additional box diagrams yielding lowest-order couplings 
between phonons and BDW fluctuations at various other wave vectors. The 
wavevector $\mathbf{p}_y$ connects the tips of the pockets at $(k_x, k_y)$ and 
$(-k_x, k_y)$,  $\tilde{\mathbf{p}}_y = \mathbf{p}_y - \mathbf{Q}_y$, 
$\tilde{\mathbf{p}}_y^\prime = -\mathbf{p}_y - \mathbf{Q}_y$, and likewise for 
$y\leftrightarrow x$.}
	\label{fig:boxmisc}
\end{figure}

Figure~\ref{fig:boxqb} shows $B_1(\mathbf{q})$ for arbitrary phonon 
momenta $\mathbf{q}$. There are no regions where it is strongly negative. 
Figure~\ref{fig:boxmisc} shows the other scattering processes that couple BDW 
fluctuations to certain phonons. The first diagram, coupling the BDW to the 
phonon with $\mathbf q = \mathbf 0$, is equivalent to $B_1(\mathbf{Q}_n)$ and 
is positive. The second and third diagrams are equal to each other and are also 
positive. All of them scale as $T^{-3/2}$ since they mainly involve the pocket 
tips.

\section{Large Fermi Surface}
\label{sec:Appendix:LargeFS}
In this appendix we discuss the influence of the Fermi surface geometry on the 
coupling between phonons and order parameter fluctuations. For the large Fermi 
surface (Fig~\ref{fig:FS}b), the leading BDW instabilities are at the diagonal 
wavevectors $(Q,\pm Q)$~\cite{Sachdev2013, Chowdhury2014a}. Consequently the 
phonon softening is strongest at these wavevectors, and our mechanism highlights 
the coupling of phonons at these wavevectors to the order parameters. The box 
diagrams coupling to the order parameters are easily computed in the linearized 
hot-spot approximation and are given by (dropping coupling constants and phonon 
form factors)
\begin{align}
&B_1(Q,\pm Q) = \tilde{B}_1(Q,\pm Q)  \sim 
\int\frac{d^2\mathbf{k}}{(2\pi)^2}T\sum_{\omega_n}\frac{1}{(i\omega_n - 
\mathbf{v}\cdot\mathbf{k})^2}\frac{1}{(i\omega_n + 
\mathbf{v}\cdot\mathbf{k})^2} =\frac{7 \Lambda_{\parallel}  \zeta (3)}{32 \pi 
^4 T^2 |\mathbf{v}|}, \\
&B_2(Q,\pm Q) = -\tilde{B}_2(Q,\pm Q) = -B_1(Q,\pm Q),
\end{align}
where $\mathbf v$ is the Fermi velocity at one hot spot. Note the exact 
cancellation between $B_2$ and $\tilde{B}_2$. The rest of the story is exactly 
the same and $(Q, Q)$-BDW order could be melted by driving the phonon with the 
same wavevector.

If the axial wavevectors $\mathbf{Q}_n$ (which is not the leading instability) 
are used, the box diagrams yield
\begin{align}
&B_1(\mathbf{Q}_n) = \tilde{B}_1(\mathbf{Q}_n)  \sim 
\int\frac{d^2\mathbf{k}}{(2\pi)^2}T\sum_{\omega_n}\frac{1}{(i\omega_n - 
\mathbf{v}_1\cdot\mathbf{k})^2}\frac{1}{(i\omega_n - 
\mathbf{v}_2\cdot\mathbf{k})^2} = 0, \\
&B_2(\mathbf{Q}_n)  \sim 
-\int\frac{d^2\mathbf{k}}{(2\pi)^2}T\sum_{\omega_n}\frac{1}{(i\omega_n - 
\mathbf{v}_1\cdot\mathbf{k})^2}\frac{1}{(i\omega_n + 
\mathbf{v}_1\cdot\mathbf{k})}\frac{1}{(i\omega_n - 
\mathbf{v}_2\cdot\mathbf{k})} = -\frac{1}{32|\mathbf{v}_1\times\mathbf{v}_2|T}, 
\\
&\tilde{B}_2(\mathbf{Q}_n)  \sim 
\int\frac{d^2\mathbf{k}}{(2\pi)^2}T\sum_{\omega_n}\frac{1}{(i\omega_n - 
\mathbf{v}_1\cdot\mathbf{k})}\frac{1}{(i\omega_n + 
\mathbf{v}_1\cdot\mathbf{k})}\frac{1}{(i\omega_n + 
\mathbf{v}_2\cdot\mathbf{k})}\frac{1}{(i\omega_n - 
\mathbf{v}_2\cdot\mathbf{k})} \nn
& \qquad \qquad = \frac{1}{16|\mathbf{v}_1\times\mathbf{v}_2|T},
\end{align}
where $\mathbf v_1$ and $\mathbf v_2$ are the Fermi velocities at the hot spots 
connected by $\mathbf Q_n$. Hence the proposed mechanism would not work because 
$B_1$ vanishes and melting of BDW correlations with wavevector $\mathbf Q_n$ is 
not possible by driving phonons. There is also no cancellation between $B_2$ and 
$\tilde{B}_2$ in this case. Having a sizable coupling between BDW fluctuations 
and phonons at wavevector $\mathbf Q_n$ thus requires a small Fermi surface.

%

\end{document}